\documentclass[journal]{IEEEtran}
\usepackage{cite}

\ifCLASSINFOpdf
\usepackage[pdftex]{graphicx}
\else
\fi
\usepackage{enumitem}

\usepackage{graphicx}
\usepackage{subcaption}
\usepackage{tcolorbox}
\usepackage{tikz}

\usepackage{url}

\usepackage{multirow}
\usepackage{float}
\restylefloat{table}
\usepackage{listings}
\usepackage{lipsum}
\usepackage{xcolor}
\usepackage{adjustbox}
\usepackage{color, colortbl}
\usepackage{array}
\usepackage{booktabs}
\usepackage{wasysym}
\usepackage{graphicx}
\usepackage{subcaption}
\usepackage{tcolorbox}
\usepackage{tikz}
\usepackage{enumitem}
\usepackage{url}

\newcommand{\mypara}[1]{\vspace{5pt}\noindent{\textit{\textbf{#1}}}}
\newcommand{\mysubpara}[2]{\vspace{2pt}{\textbf{C#1}} {\textit{#2}}}
\def\textsc#1{\textnormal{{\sc #1}}}%

\begin{document}
%
\title{\textsc{uTango}: an open-source TEE for IoT devices}
%
%
%
\author{
\IEEEauthorblockN{Daniel Oliveira,
Tiago Gomes, 
Sandro Pinto}
\IEEEauthorblockA{\\Centro ALGORITMI - University of Minho\\
\{daniel.oliveira, mr.gomes, sandro.pinto\}@dei.uminho.pt}}

\maketitle

\begin{abstract}
  Security is one of the main challenges of the Internet of Things (IoT). IoT devices are mainly powered by low-cost microcontrollers (MCUs) that typically lack basic hardware security mechanisms to separate security-critical applications from less
  critical components. Recently, Arm has started to release Cortex-M MCUs enhanced with TrustZone technology (i.e., TrustZone-M), a system-wide security solution aiming at providing robust protection for IoT devices. Trusted Execution Environments (TEEs) relying on TrustZone hardware have been perceived as safe havens for securing mobile devices. However, for the past few years, considerable effort has gone into unveiling hundreds of vulnerabilities and proposing a collection of relevant defense techniques to address several issues. While new TEE solutions built on TrustZone-M start flourishing, the lessons gathered from the research community appear to be falling short, as these new systems are trapping into the same pitfalls of the past. 
  In this paper, we present \textsc{uTango}, the first multi-world TEE for modern IoT devices. \textsc{uTango} proposes a novel architecture aiming at tackling the major architectural deficiencies currently affecting TrustZone(-M)-assisted TEEs. In particular, we leverage the very same TrustZone hardware primitives used by dual-world implementations to create multiple and  equally secure execution environments within the normal world. We demonstrate the benefits of \textsc{uTango} by conducting an extensive evaluation on a real TrustZone-M hardware platform, i.e., Arm Musca-B1. \textsc{uTango} will be open-sourced and freely available on GitHub in hopes of engaging academia and industry on securing the foreseeable trillion IoT devices.
\end{abstract}


\bstctlcite{IEEEexample:BSTcontrol}

\section{Introduction}



With the increasing complexity of the Internet of Things (IoT) devices and the door left open by Internet connectivity to hackers and attackers, developing secure IoT devices is becoming increasingly challenging \cite{Keoh2014, Alrawi2019, Balzarotti2020}. Complex functional requirements are met by integrating multiple codebases, drivers, and libraries from different 3rd party entities with distinct assurance levels. Such problems are exacerbated in devices composed by microcontrollers (MCUs) that typically lack reliable mechanisms to enforce the separation among these multi-source and mixed-criticality components \cite{Pan2018, Pinto2019_Voila}. Therefore, with the IoT ecosystem evolving at a breathtaking pace, the need for reliable security architectures is rising; however, due to the heterogeneity and limitation of computational resources on MCUs, creating and adopting these architectures still challenges most IoT system developers \cite{Sadeghi2015,Oliveira2020,Luo2020,Freiling2017}.
 

In the context of secure computing systems, security through separation is a well-established principle implemented by microkernels, hypervisors, and Trusted Execution Environments (TEEs) \cite{Klein2009, Koeberl2014, Brasser2015, Sun2015,  Costan2016,  Ferraiuolo2017, Maene2018, Hassaan2019, Brasser2019, Pinto2019_Survey, Li2019, Hahm2020, Lee2020, Bahmani2021}. In particular, billions of mobile devices operating worldwide rely on TEEs leveraging TrustZone hardware primitives for the protection of security-critical applications (e.g., digital rights management, fingerprints, and keys) \cite{Sun2015, Hua2017, Ferraiuolo2017, Pinto2019_Survey, Brasser2019, Pinto2020}.
TrustZone enables the partition of system resources into two domains: the \textit{secure world} for secrets and critical functionality; and the \textit{normal world} for everything else, including the operating system (OS) and its applications. Within the realm of the tiniest IoT devices, Arm has adapted TrustZone technology for the Cortex-M family, introducing TrustZone-M into the new Armv8-M MCUs, e.g., Cortex-M23 and Cortex-M33 \cite{Nyman2017, Asokan2018, Pinto2019_Voila}.

\mypara{Problem and Motivation.}
TrustZone-assisted TEEs are assumed to be highly secure. However, over the past years, TrustZone-assisted TEEs have been attacked several times \cite{Zhang2016,  Tang2017, Pinto2019_Survey, Keegan2019, Cerdeira2020}. A recent study performed on five major commercial TrustZone-assisted TEEs (i.e., Qualcomm, Trustonic, Huawei, Nvidia, and Linaro) has identified that TrustZone-assisted TEEs have several \textit{architectural deficiencies}, critical \textit{implementation bugs}, and overlooked \textit{hardware properties} \cite{Cerdeira2020}. The architectural issues have the largest share from these three classes of problems. Among the identified deficiencies, we highlight (i) the excessively large trusted computing base (TCB), e.g., QSEE has 1.6 MiB, (ii) a large number of interfaces, e.g., QSEE has 69 syscalls, (iii) the existence of several privileged secure kernel drivers, and (iv) the asymmetrical isolation between the worlds, e.g., trusted applications can map normal world memory \cite{Cerdeira2020}. 

In general, although these problems have mainly affected commercial TEE systems targeting Cortex-A processors, new TEE solutions built on TrustZone-M MCUs are falling into the same pitfalls of the past. The Arm Trusted Firmware-M (ATF-M) \cite{atfm2020} implements a large number of kernel components and security services within the secure world and existing memory protection mechanisms (i.e., secure Memory Protection Unit (MPU)) are configured with too coarse-grained regions. As a result, ATF-M has a TCB with hundreds of KiloBytes (150+ KiB). The Kinibi-M was adapted from the original Kinibi TEE for mobiles and not re-invented for the IoT \cite{kinibi2021}. And lastly, Arm is currently spreading an ambiguous message concerning what should be deployed within the secure world. Multiple official Arm TrustZone-M documents \cite{Arm2018, Arm2019, ArmKeil2019} suggest different approaches. Very critical, in Ref. \cite{Arm2018}, for an IoT application targeting a wireless communication interface, Arm suggests including (i) the secure boot, (ii) the communication stack, (iii) device drivers, (iv) OS kernel, and (v) firmware update within the secure world and a communication buffer on the normal world. 

\mypara{Existing Solutions.} As the current dual-world model is becoming inadequate to address the increasing complexity and requirements of modern IoT devices, the academia presented several works that extend this model into unique multi-domain TEE architectures aiming to address this critical limitation. TrustICE \cite{Sun2015} and Sanctuary \cite{Brasser2015} are two prominent solutions that extend the TrustZone model by creating full-isolated enclaves within the normal world. In another line of works, the vTZ \cite{Hua2017}, OSP \cite{Cho2016}, and PrivateZone \cite{Jang2018} solutions leverage virtualization mechanisms to provide multiple isolated environments within the normal world. However, all these works are developed on top of high- to middle-end processors, which feature mechanisms not available on MCUs. Targetting low-end MCUs, the number of systems that propose architectures with multiple isolated environments is still scarce. As of this writing, the commercial Multizone TEE \cite{Garlati2020} is one solution that follows this multi-domain concept and distinguishes it from other traditional systems such as ATF-M, Kinibi-M, and ProvenCore-M. To the best of our knowledge, the Multizone TEE is the closest solution to \textsc{uTango}, but it targets Armv7-M processors, which inherently brings several disadvantages, namely, the need for trap and emulation techniques and binary translation of privileged instructions. Thus, we believe that a novel multi-world architecture enabling the execution of multiple environments within strongly isolated compartments would provide higher flexibility and increase security guarantees for the new generation of MCU-powered IoT devices.

\mypara{Contributions.} In this paper, we present \textsc{uTango}, an open-source TEE that aims at tackling the main architectural flaws that currently affect TrustZone(-M)-assisted TEEs (i.e., large TCBs and poor isolation between different world domains \cite{Cerdeira2020}). To shrink the TCB, \textsc{uTango} sits solely in the secure world and creates an unlimited number of non-secure domains to host not only normal applications but also typical TEE-based kernel components (e.g., secure services). To ensure the strict isolation of virtual worlds, \textsc{uTango} leverages the dynamic reconfiguration capabilities of TrustZone-M controllers. Each controller is dynamically programmed to partition system resources according to each execution environment memory, devices, and interrupt assignments. To the best of our knowledge, \textsc{uTango} is the first multi-world TEE for TrustZone-M IoT devices. The novel TEE architecture leverages the very same TrustZone-M hardware primitives used by dual-world implementations to provide multiple equally-secure execution environments and augmented TEE capabilities (e.g., more than two sandboxed environments and availability guarantees).

The design follows three main principles: (i) \textit{principle of minimal implementation}, by providing a minimal and clean-slate implementation, with a small number of well-defined interfaces, thereby drastically reducing the overall system TCB; (ii) \textit{principle of least privilege}, by ensuring that \textsc{uTango}, as the highest privilege entity, is the single component running within the secure world; (iii) and \textit{principle of containment}, by enforcing that each functional block executes on its isolated execution domain, which inherently prevents lateral movement and privilege escalation. 
To comply with these design principles, the materialization of \textsc{uTango} comes with several interesting challenges to be resolved: (i) how to enforce strict separation between worlds leveraging adequate usage of typical seldom MCU-based protection mechanisms; (ii) how de-privileged security services need arbitrated access to hardwired secure peripherals; (iii) how \textsc{uTango} must provide a set of secure communication channels to worlds exchange messages between them; and (iv) how to offer tangible improvements in real-world scenarios, i.e., \textsc{uTango} must be as less intrusive as possible and provide near-native performance.

\par We have implemented \textsc{uTango} on the Arm Musca-B1 platform, where we performed an extensive performance evaluation by using the Embench benchmark suite. Our comparison between native runs of the benchmark against different configurations of the benchmark under the \textsc{uTango} system shows that \textsc{uTango} introduces a minimal performance overhead, averaging - in a minimal configuration of the system - a residual value of 0.05\%. As the number of environments increases, we observed that the performance overhead increases linearly, which we conclude to be a natural result of \textsc{uTango}'s design. Moreover, the TCB of the system ($\approx$ 4.3 KiB) is an order of magnitude smaller than alternative solutions. 



\par Finally, as TrustZone's dual-world model has proved its inability to cope with the increasing complexity of modern devices, new solutions that enable multiple isolated execution environments have arisen to be suitable architectures \cite{Brasser2019, Cerdeira2020, Jang2018, Hua2017, Garlati2020, Sun2015, Cho2016, Hua2017}. With \textsc{uTango} TEE, our main goal is to tackle the aforementioned problems and emphasize that multi-world architectures are possible to endow orthogonally in all TrustZone systems, specifically in TrustZone-M-based devices that bring another set of challenges. To that end, our main contributions are as follows: 

\begin{enumerate}[leftmargin=*]

    \item We present the design of \textsc{uTango} as a novel TEE architecture leveraging TrustZone-M hardware primitives to provide an unlimited number of equally-secure execution environments  (Section \ref{ud}). 
    
    \item We provide a proof-of-concept implementation of \textsc{uTango} targeting the first public available TrustZone-M platform, i.e., Arm Musca-B1 (Section \ref{ui}). All software components will be open-sourced and available on GitHub.
    
    \item We perform a comprehensive security analysis and discuss how \textsc{uTango} can mitigate potential attack vectors that a malicious adversary may explore (Section \ref{sa}). 
    
    \item We extensively evaluate \textsc{uTango} focusing on security metrics, performance, interrupt latency, and TCB size and overall complexity (Section \ref{ev}). Our results demonstrate why our solution has minimal impact on the overall system and why it is a perfect fit for modern resource-constrained IoT devices. 
    
\end{enumerate}


\section{Background}

\subsection{Arm TrustZone-M}
\label{tz-m}

\par Arm TrustZone follows a system-wide approach to security, providing hardware-enforced protection mechanisms at the CPU and System-on-Chip (SoC) level \cite{Pinto2019_Survey}. This technology is centered around the concept of protection domains named \textit{secure world} and \textit{normal world}. TrustZone was firstly introduced into Arm application processors (Cortex-A) in 2004, achieving mainstream adoption in the mobile industry. 
The technology has provided the hardware foundations to foster the creation of mainly TEE systems \cite{Sun2015,  Hua2017, Ferraiuolo2017, Pinto2019_Survey, Brasser2019, Pinto2020}, but also virtualization infrastructures \cite{Pinto2019_Voila, Pinto2017}. However, the later solutions still face several limitations, and, therefore, the technology has been continuously being more used in the context of enabling TEEs.  
\par In 2016, Arm decided to span TrustZone for the new generation of Arm MCUs, i.e., Armv8-M (e.g., Cortex-M23 and Cortex-M33), naming this new version of the technology as TrustZone-M. From a high-level perspective, both technologies follow the same dual-world architecture. However, at the low level, there are significant differences, mainly because TrustZone-M is entirely optimized for low-end devices (e.g., deterministic behavior, low overhead, and low-power consumption) \cite{Arm2018}. 

\label{bg1}
\mypara{Programming Model.} Armv7-M MCUs provide two operation modes: \textit{thread} and \textit{handler} mode. In thread mode, the processor executes application code, which can be either privileged or unprivileged. In handler mode, the processor executes exception handler code, which is always privileged. In (Armv8-M) TrustZone-enabled MCUs, these operation modes are orthogonal to the two security states, i.e., there is both a thread and handler mode for each security state. The security state does not depend on a specific security bit, but the division is memory map based. This means that when the code is running from the secure memory, the processor state is secure, and when the code runs from non-secure memory, the processor state is non-secure. Transitions between the two worlds are supported by three new instructions: branch with exchange to non-secure state (BXNS), secure gateway (SG), and branch with link and exchange to non-secure state (BLXNS). Calling non-secure software from the secure state is possible by performing a BLXNS instruction. In contrast, non-secure software cannot directly call secure software. Instead, non-secure software must use indirect entry points located in a Non-Secure Callable (NSC) memory region. The first instruction of any entry point must be an SG, which marks a valid branch to secure code. After the secure function completes, a BXNS instruction issues a return to the non-secure software. Furthermore, state transitions can also happen due to exceptions or interrupts. 

\label{bg2}
\mypara{Resources Partitioning.}
In Armv8-M MCUs, the memory space is partitioned through the so-called attribution units. The Security Attribution Unit (SAU) is always available and provides dynamic address partitioning. The chip designer defines the number of regions, which is typically 8. The SAU is programmable in the secure state. The Implementation-Defined Attribution Unit (IDAU) is external to the core and implementation-defined. The IDAU provides static address partitioning and supports up to 256 non-programmable regions. The configuration of the memory region's security state results from the $OR$ logic operation between the SAU and IDAU. Memory can also have a set of privilege permissions defined by a TrustZone-aware Memory Protection Unit (MPU). MPUs are banked among worlds; however, MPUs are optional and implementation-defined.
Additionally to the attribution units, other components, referred to as \textit{security wrappers} and \textit{gates} (i.e., block-based, watermark-based, and select-based), propagate the security model defined at the core level to the remaining system bus masters and slaves. Security wrappers are placed in front of non-security aware masters to wrap their transactions into secure or non-secure transactions. Flash and SRAM memory slaves can use block-based (memory divided into multiple, alternating blocks of secure and non-secure regions) or watermark-based gates (memory splitted into two regions, one secure and the other non-secure) to filter transactions. The select-based gates are used to filter transactions based on the device slave address and the assigned world. These components are controlled by a central system security controller, speciﬁed by silicon manufacturers.

\label{bg3}
\mypara{Interrupt Handling.} In TrustZone-enabled MCUs, interrupts can be set as secure or non-secure by configuring the Interrupt Target Non-secure (ITNS) interface on the Nested Vector Interrupt Control (NVIC). Arm's M-profile architectures support automatic hardware stacking and un-stacking of some CPU registers during exception entrance to reduce the interrupt latency. Armv8-M-based architectures follow the same concept, with notable exceptions. If the arriving interrupt has the same security state as the processor, the execution flow is almost identical. The main difference occurs when a non-secure interrupt triggers while secure code is executing. To avoid information leakage, the processor automatically pushes all non-banked registers to the secure stack and erases all its contents, which increases interrupt latency. The vector table is banked between states, i.e., the processor supports two separated exception vector tables. Furthermore, secure and non-secure interrupts can share the same priority level, or secure interrupts can be programmed to have higher priority than non-secure ones (i.e., to avoid denial-of-service (DoS) attacks). 

\label{bg4}
\mypara{Findings of Security Observations.} Arm TrustZone-M is already subject to some security issues and disclosed vulnerabilities. Luo et al. \cite{Luo2020} presented a comprehensive security analysis that observed several potential software security issues in TrustZone-M-enabled MCUs. The list of potential exploits to subvert TrustZone-M's security foundations are based on four types of attacks: (i) code injection, (ii) code reuse, (iii) heap-based buffer overflows, (iv) format strings, and (v) specific NSC attacks. These attacks were successfully performed in the Microchip SAML11 platform with TrustZone-M extensions. Arm has also unveiled another high impact vulnerability (CVE-2020-16273) that enables a non-secure adversary to manipulate the secure world control flow \cite{Arm2020}. The attack exploits the poor management of the secure stacks that can open doors to a malicious attack that causes incorrect code execution through a stack underflow scenario. 


\subsection{Arm Trusted Firmware-M}
\label{atf-m}


\par The Arm Trusted Firmware-M (ATF-M) is an open-source, secure world firmware reference implementation, which offers the foundations of a TEE for Armv8-M MCUs. The ATF-M implements: (i) a secure boot to verify the integrity and authenticity of secure and non-secure binaries; (ii) a core module (i.e., ATF-M Core) that controls the isolation, communication, and execution; and (iii) a set of security services offering secure storage, crypto, and attestation mechanisms.
The implementation follows the traditional TrustZone dual-world architecture, i.e., all secure components (e.g., bootloader, kernel modules, secure services, and 3rd party security functions) are encapsulated within the secure world. The solution implements the isolation levels defined in the Platform Security Architecture (PSA) \cite{armpsa2021}, which rely on platform hardware (e.g., SAU, secure MPU, etc.) to enforce isolation boundaries. As of this writing, ATF-M only implements isolation levels 1 and 2, which partitions the system into three major domains. Isolation level 1 establishes the two specific security domains enabled by TrustZone (i.e., the secure and non-secure worlds). Isolation level 2 goes a step further and leverages the secure MPU to isolate the ATF-M core and services from 3rd party security services. These services are expected to be developed by multiple entities, and are not the same services as the ones provided by the ATF (i.e., secure storage, attestation, etc.). Isolation level 3 is still under development and will provide fine-grained isolation among different 3rd party security services.

%

\begin{table}[t!]
\centering
\caption{ATF-M TCB size, code size, and security metrics for NUCLEO-L552ZE-Q and Musca-B1 platforms.}
\label{atf-table}
\scalebox{0.80}{
\begin{tabular}{ p{2.4cm} || p{2.0cm}  p{1.3cm}  p{1.4cm} }

\multicolumn{1}{c}{\textbf{}} & & \cellcolor{lightgray!50}\multirow{1}{1.4cm}{\centering\textbf{\small STM32L5}}  & \multirow{1}{1.4cm}{\centering\textbf{\small \cellcolor{lightgray!50}Musca-B1}} \\


\hline
\multirow{3}{2.4cm}{\centering\textbf{\small Binary Size (KiB)}}  & \multirow{1}{*}{\small \textit{bootloader}}       & \multicolumn{1}{c}{\small 103} & \multicolumn{1}{c}{\small 50} \\
                                                                & \multirow{1}{*}{\small \textit{core+services}}    & \multicolumn{1}{c}{\small 192} & \multicolumn{1}{c}{\small250} \\
                                                                & \multirow{1}{*}{\small \textit{total}}            & \multicolumn{1}{c}{\small 295} & \multicolumn{1}{c}{\small300} \\
\hline

\multirow{1}{2.5cm}{\centering\textbf{\small Code Size (SLoC)}}   & \multirow{1}{*}{\small \textit{total}}            & \multicolumn{1}{c}{\small 33 K}  & \multicolumn{1}{c}{\small 40 K} \\

\hline

\multirow{2}{2.4cm}{\centering\textbf{\small Security Metrics}}   & \multirow{1}{*}{\small \textit{\# ROP Gadgets}}   & \multicolumn{1}{c}{\small 9957} & \multicolumn{1}{c}{\small 15597} \\
                                                                & \multirow{1}{*}{\small \textit{\# Indirect Calls}}& \multicolumn{1}{c}{\small 147} & \multicolumn{1}{c}{\small 257} \\

\hline
\end{tabular}
}
\end{table}

\mypara{ATF-M preliminary evaluation.}
To understand the ATF-M codebase complexity, we performed a preliminary evaluation with regard to binary size, code size, and security metrics of the ATF-M implementation for two different platforms: Arm Musca-B1 and STM NUCLEO-L552ZE-Q. Table \ref{atf-table} summarizes the assessed results.
%
%
The off-the-shelf implementation of the ATF-M for the Arm Musca-B1 has a total size of 50 KiB for the secure bootloader and 250 KiB for the ATF-M core and security services. The implementation has a large code size, encompassing approximately 40 K source lines of code (SLoC). We also evaluated the number of indirect calls and Return-Oriented Programming (ROP) gadgets, which are important security metrics as further explained in Section \ref{ev}. We counted 257 indirect calls and 15597 ROP gadgets across the different binaries that comprise the TCB. The numbers for the STM port are slightly better but in the same order of magnitude, i.e., 192 KiB for the ATF-M core and security services, 33 K SLoC, 9957 ROP gadgets, and 147 indirect calls. 
In summary, this preliminary evaluation suggests that the same architectural issues highlighted in the literature \cite{Cerdeira2020} are being repeated. Furthermore, the formally verified microkernel seL4 \cite{Klein2009}, which targets high-end Linux-capable platforms endowed with a memory management unit (MMU), has a smaller TCB and a simpler codebase \cite{Cerdeira2020} than ATF-M, that is expected to run on resource-constrained MCUs.



\mypara{Findings of Security Vulnerabilities.} 
Since its release, ATF-M has already provided some fixes to solve software-related security vulnerabilities, namely CVE-2021-27562, CVE-2021-32032, and CVE-2021-40327 \cite{atfm2022}. The first security issue directly affected the inter-process communication model, where an attacker could potentially crash the secure world or reset the whole system. Regarding the other two vulnerabilities, both affect the Crypto secure service and cause memory leakage that corrupts the service or even leak the secure keys to the non-secure world. Such findings need to be carefully addressed, and as the solution starts to spread, more vulnerabilities are likely to surface as the increasing TCB further opens the attack surface of ATF-M.

\begin{figure}[t]
    \centering
    \begin{subfigure}[b]{0.47\textwidth}        
        \centering
        \includegraphics[width=0.8\linewidth]{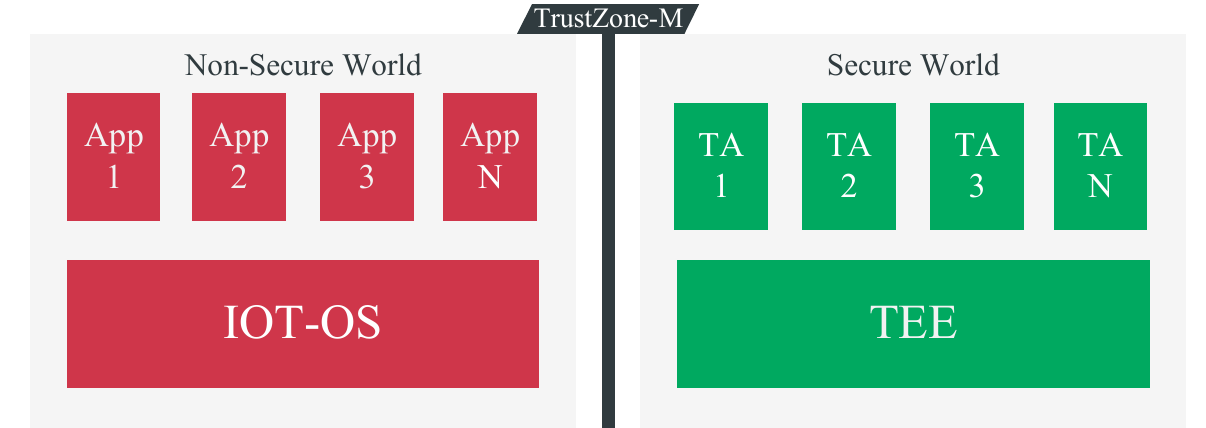}
        \caption{Classic TrustZone-assisted TEE architecture with a dual-world model. Secure applications sit on the secure world, while everything else is placed in the non-secure world.}
        \label{fig:design1}
    \end{subfigure}
    \begin{subfigure}[b]{0.47\textwidth}        
        \centering
        \includegraphics[width=1\linewidth]{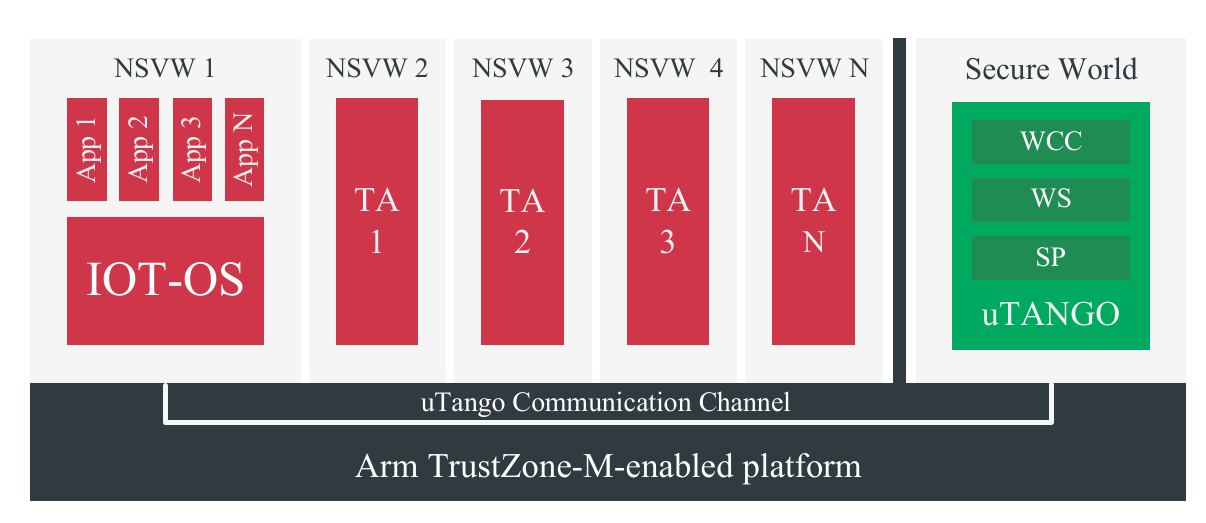}
        \caption{uTango TEE architecture with a novel multi-world model. Compared to a), solely uTango sits on the secure world, while everything else (including trusted applications) is assigned and isolated in an NSVW.}
        \label{fig:design2}
    \end{subfigure}
    \caption{Architectural differences between \textsc{uTango} and typical TEE systems.}
    \label{fig:design}
\end{figure}

\section{uTango Design}
\label{ud}

\par \textsc{uTango} aims at tackling the main architectural deficiencies prevailing on TrustZone-(M)-assisted TEE systems \cite{Cerdeira2020}. To do so, \textsc{uTango} evolves from the classic dual-world security model (Figure \ref{fig:design1}) to a \textit{multi-world} architecture (Figure \ref{fig:design2}). The multi-world architecture is based on the zero-trust model, which dictates that every single software component, with the exception of the TEE kernel, cannot be trusted. Thus, \textsc{uTango} enables the consolidation of multiple applications, services, or workloads (e.g., embedded OSes) on equally secure, isolated domains - called \textit{Non-Secure Virtual Worlds} (NSVW). 


Another particularity of the \textsc{uTango} design, is related to the augmentation of the TEE model and capabilities. TrustZone-(M) TEEs implement remote procedure call (RPC) architectures, i.e., a client-server model mainly used for mobiles. Modern TEEs aim at providing a broader range of features and fulfilling a much large spectrum of use cases and requirements \cite{Lee2020, Garlati2020}. Thus, \textsc{uTango} not only supports the traditional core of the TrustZone design but also extends it with more than two fully-sandboxed worlds and offers availability guarantees \cite{Sun2015, Brasser2019, Li2019}. Within the NSVWs, \textsc{uTango} runs unmodified binaries, which can be user-space applications, services, libraries, or privileged OS/RTOS and respective applications. Furthermore, \textsc{uTango} goes a step further by providing increasing availability guarantees.





\mypara{Design Principles.} 

The design of the \textsc{uTango} multi-world architecture is centered on three fundamental principles, i.e., \textit{principle of minimal implementation}, \textit{principle of least privilege}, and \textit{principle of containment}, which are critical during the design of (secure) TEE systems \cite{Saltzer1975}. Such principles are inherently followed by other systems that enable multiple isolated execution environments in their infrastructures \cite{Brasser2019, Jang2018, Hua2017, Sun2015, Cho2016, Hua2017}. Therefore, \textsc{uTango} complies with them, and we show that its systematic application results in (i) a reduced TCB and attack surface, (ii) a well-defined layered access control to prevent privilege escalation, and (iii) strong isolation boundaries to restrict workloads access to only their resources (preventing exploits containment).


\begin{tcolorbox}[title=Principle of minimal implementation]
To contain the system's attack surface, \textsc{uTango} must rely on hardware support as much as possible, and provide a minimal and clean implementation of well-defined structures and interfaces. Moreover, \textsc{uTango} must de-privilege secure applications/services to the normal world, thus reducing the amount of code to be trusted and deployed on the secure side. 
\end{tcolorbox}

\begin{tcolorbox}[title=Principle of least privilege]
To mitigate privilege escalation, \textsc{uTango} kernel must be granted the highest privilege of execution while de-privileging secure applications/services to the normal world. Furthermore, each execution domain must only have access to those resources that are absolutely required (e.g., devices and system services).  
\end{tcolorbox}

\begin{tcolorbox}[title=Principle of containment]
To limit the extent of an attack, \textsc{uTango} must ensure that the multiple execution domains are well-defined and self-contained with clear boundaries. The system must use hardware-enforced mechanisms to sandbox each domain to its resources (e.g., memory, devices, and interrupts), thereby limiting the reach of an attacker and preventing lateral movement across other system components. 
\end{tcolorbox}


\subsection{Architecture Overview}
\label{ud1}

\par Figure \ref{fig:design1} depicts the traditional dual-world TrustZone-M architecture. The \textsc{uTango} architecture, illustrated in Figure \ref{fig:design2}, presents a multi-world scheme, where each functional block (i.e., IoT-OS, baremetal, or trusted applications) is mapped to an NSVW and sandboxed under its boundaries. Thus, as highlighted in Figure \ref{fig:design2}, the \textsc{uTango} kernel is the single component running at the most privileged level (i.e., secure handler mode), while all NSVWs run in the non-secure state. Following the aforementioned design principles, \textsc{uTango} must strive for a clean-slate minimal implementation. Thereby, the \textsc{uTango} kernel is built around three components: (i) the system partitioner (SP), (ii) the worlds scheduler (WS),  and (iii) the worlds communication channel (WCC). The SP relies on a configuration file detailing the overall system configuration and partition. 

\mypara{System Configuration.} 
The first piece of \textsc{uTango} workflow starts with a configuration file that defines each NSVW's properties. These properties encompass memory regions (e.g., code and data), devices (e.g., serial peripherals, timers, etc.), interrupts assigned to each NSVW, and the overall system time quantum.
 

\begin{figure*}[ht]
    \centering
    \includegraphics[width=0.9\textwidth]{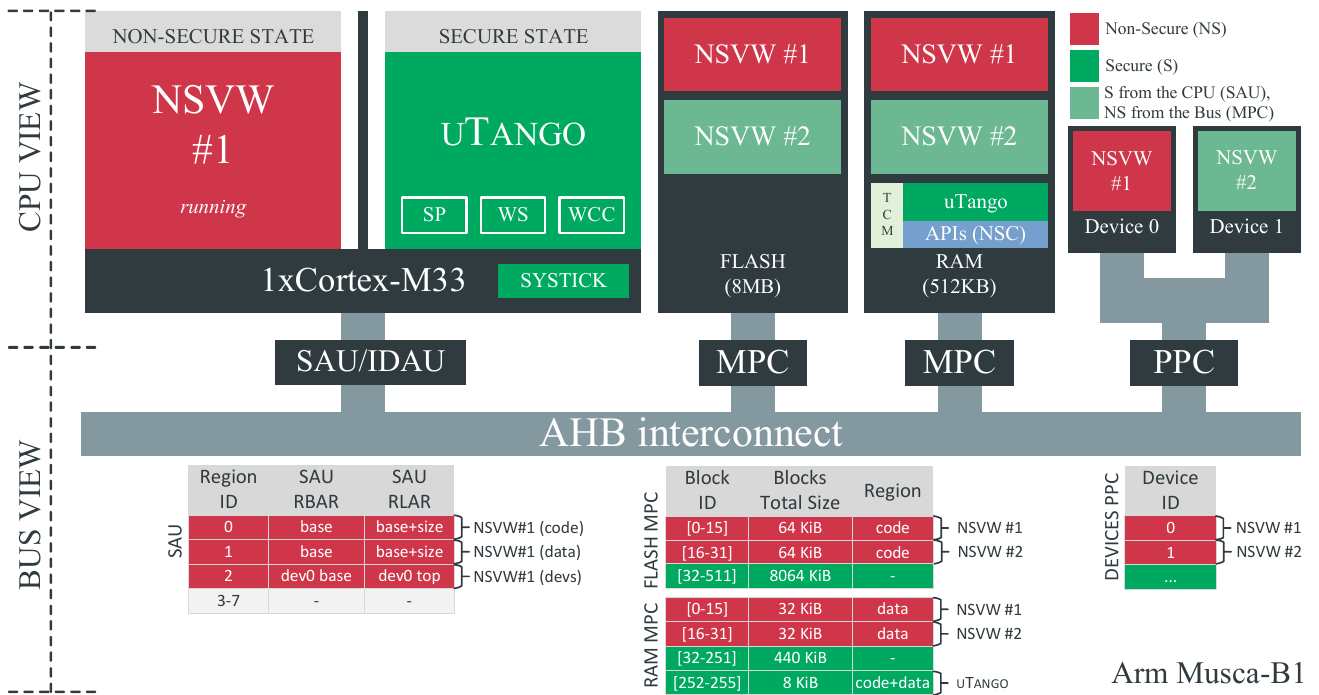}
    \caption{\textsc{uTango} full system implementation view (CPU and bus) for the Arm Musca-B1 targeting a two NSVWs configuration. Snapshot of system configuration while NSVW \#1 is running.}
    \label{fig:impl}
\end{figure*}

\mypara{System Partitioner.} 
The SP is responsible for partitioning the platform resources at boot-time according to the system configuration. The SP leverages the SAU and additional platform-specific bus filters (security gates) to achieve such partitioning. The number of SAU regions limits only the number of memory-mapped resources per NSVW and not the number of NSVWs. The maximum number of NSVWs is only limited by the amount of available memory in the target platform. Based on each NSVW configuration, the SP prepares a corresponding SAU configuration and saves it on the world's control block (WCB). Regarding the security gates, the SP only performs a one-time setup. \textsc{uTango} enforces that all accesses and transactions issued by NSVWs to other bus masters (e.g., DMA, cryptographic engines, etc.) are always trapped and mediated. This prevents the reconfiguration of each bus filter during the context switch, avoiding additional performance burdens. 



\mypara{Worlds Scheduler.} 
\textsc{uTango} enforces temporal separation through the WS. According to the system time quantum, the WS, supported by an architectural timing unit (e.g., Arm SysTick), schedules each NSVW in a round-robin fashion. Every NSVW has a unique WCB data structure for preserving the world state, i.e., CPU register bank, selective System Control Block (SCB) registers, SAU configuration table, and interrupts state. At every scheduling point, the WS performs four main activities: (i) saves the suspended NSVW context to the WCB; (ii) schedules a new NSVW to be resumed; (iii) sets new partition regions on SAU; and (iv) restores the context of the new NSVW. Notice that, while setting a new configuration to the SAU, the resources from suspended NSVWs are preserved and marked as secure, preventing possible unauthorized accesses from the running NSVW.

\mypara{Worlds Communication Channel.} 
The \textsc{uTango} offers a communication infrastructure that allows the exchange of secure messages across NSVWs. 
To reduce the risk of vulnerabilities, \textsc{uTango} uses a message-passing mechanism, i.e., no-shared memory. 
Shared memory channels across isolated boundaries are perceived as a significant source of vulnerability in secure systems since its data must be treated as untrusted and volatile \cite{ArmPSA2019}.
Messages have a fixed 12-byte data stream length. \textsc{uTango} provides four APIs (i.e., blocking and non-blocking) to send and receive messages. The WCC acts as a messaging gateway, checking each NSVW's inbox and forwarding each message to its respective NSVW. The system designer is responsible for defining messages and semantic for the application. Standardized APIs, such as the Global Platform TEE API, can be built atop these primitives.



\section{uTango Implementation}
\label{ui}

\mypara{System Setup.} \label{ui1} 
The \textsc{uTango} TEE was firstly implemented for the Arm Musca-B1 Test Chip Board, which implements the SSE-200 subsystem that features a multi-core system with two Cortex-M33 processors. Despite the dual-core architecture, \textsc{uTango} currently only supports a single-core configuration. 
Figure \ref{fig:impl} depicts a low-level view of \textsc{uTango} running a two NSVW configuration on the Arm Musca-B1. At the time of this writing, we are currently working on the support for two additional platforms: NXP LPC55S69-EVK and STM NUCLEO-L552ZE-Q. 


\mypara{uTango Hardware Components.} \label{ui2} 
The main software components of \textsc{uTango} rely on hardware primitives available on TrustZone-based MCUs. As previously mentioned, the WS uses the secure SysTick as the temporal source for scheduling all NSVWs. To partition the system, the SP configures the SAU to overlap the fixed IDAU memory security regions and specify the overall system memory layout. With the SAU correctly configured, core transactions (including data read/write, instruction fetches, and debug access) are secured. As previously stated, we assume bus masters are always secured and managed by \textsc{uTango}, so TrustZone-aware bus slaves (i.e., memories and peripherals) need to be configured according to the overall system security model. 
We configure Musca-B1 security gates, i.e., the block-based Memory Protection Controllers (MPC) and the select-based Peripheral Protection Controller (PPC), to match all NSVWs' memory and devices assignments. The remaining memory blocks or device sets are kept secure.

\mypara{uTango Secure Boot.} \label{ev-sa4}
The processor starts in the secure state by default, enabling root-of-trust implementations such as the secure boot. \textsc{uTango} features a 2-stage secure bootloader to verify the integrity and authenticity of the firmware image against a store signature (SHA-512) held in secure memory. 
Upon detecting a verification error, the \textsc{uTango} boot is aborted, and the system is locked in a secure state until the next reset/power cycle. If the image is valid, the boot process continues until it handles control to the first NSVW to run.



\subsection{Execution Life Cycle} 
\label{ui3}

\par
\begin{figure*}[ht]
    \centering
    \includegraphics[width=0.95\textwidth]{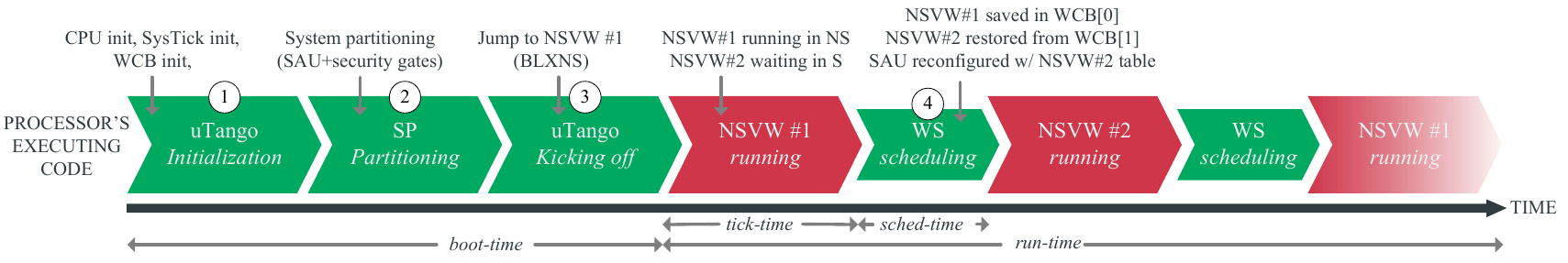}
    \caption{\textsc{uTango} execution life cycle for a system configured with 2 NSVWs.}
    \label{fig:lifecycle}
\end{figure*}

 The \textsc{uTango} execution life cycle for a system configured with two NSVWs is depicted in Figure \ref{fig:lifecycle}. The complete boot process consists of three stages: (i) \textit{Initialization}; (ii) \textit{SP Partitioning}; (iii) and finally \textit{Kicking-off}. At run-time, \textsc{uTango} is just responsible for the \textit{WS Scheduling}. In the following, we describe each stage.

\mypara{\textcircled{1} {uTango} Initialization.} \label{ui4}
After reset, the \textsc{uTango} boot agent initializes preliminary CPU- and platform-specific hardware components. Then, it reads the full system raw binary file (loaded onto the Flash) and copies each software piece (i.e., \textsc{uTango} kernel and NSVWs) to its respective memory region. The boot agent is also responsible for verifying the integrity and authenticity of the \textsc{uTango} kernel image and, in case of success, copying the image to the Tightly-Coupled Memory (TCM). This implementation detail ensures almost all \textsc{uTango} kernel instructions and memory accesses take 1-2 clock cycles because (i) there are no wait states and bus/memory stalls (performance) and (ii) code and data are not cached (security). After copying each NSVW to the respective memory segment, the initialization concludes by configuring the secure MPU to enforce policies among TEE kernel code and data sections, setting up the vector table address of \textsc{uTango}, and jumping to the main initialization routine. 
After boot, the \textsc{uTango} kernel starts executing by first enabling and configuring fault exceptions. Non-secure exceptions are configured with lower priority than secure ones, thus preventing starvation of the secure side, i.e., avoiding DoS attacks. The secure SysTick timer is then configured according to the system quantum configuration, i.e., the reload value is loaded, and the timer exception is enabled. 
The last part of the \textit{Initialization} process fills the internal WCB structures with the respective NSVWs' static configurations. The WCB encompasses 11 general-purpose registers (\textit{r4-r14}), 8 special purpose registers (i.e., \textit{msp}, \textit{psp}, \textit{msp\_lim}, \textit{psp\_lim}, \textit{basepri}, \textit{primask}, \textit{faultmask}, and \textit{control}), and a subset of specific SCB registers (e.g., \textit{vtor}, \textit{scr}, etc.). To speed up the re-configuration of the world switching operation, the SAU configuration for each NSVW is defined as part of the WCB. The last part of the WCB includes an interrupt descriptor that keeps the NVIC registers' context (e.g., priority level, enable and pending status, and security state), which we further detail in Section \ref{ui8}. 

\mypara{\textcircled{2} SP Partitioning.} \label{ui5}
After the \mbox{\textit{\textsc{uTango} Initialization}}, the execution flow continues through the \mbox{\textit{SP Partitioning}}. The SP is responsible for leveraging all available hardware mechanisms to partition the system resources according to the NSVW's settings. First, the SP unrolls all NSVW's memory regions and checks for overlapping regions; if an overlap is identified, \textsc{uTango} aborts execution. If all memory regions are valid, the SP starts programming the SAU with the memory regions assigned to the first NSVW (NSVW \#1 in Figure \ref{fig:impl}). The SAU is programmed through a set of memory-mapped registers, namely the Region Number Register (RNR), Region Base Address Register (RBAR), and Region Limit Address Register (RLAR). The RNR controls which region is active or selected, while the RBAR and RLAR set its start address and limit, respectively. This process results in a configuration similar to the SAU table depicted in Figure \ref{fig:impl}. 
After programming the SAU, the SP uses the platform-specific memory gates, i.e., the MPCs, to configure the system's overall memory partition. The MPC is a block-based gate that divides the memory into multiple, alternating blocks of secure and non-secure regions. Transactions are filtered based on the programmed regions. Each block has a well-defined size, which can be configured as secure or non-secure. Therefore, the SP must first check if a set of blocks can represent all NSVW's memory regions. If regions are within the bounds of the MPCs, the SP configures the controller, which results in a configuration similar to the Flash and RAM MPC's tables depicted in Figure \ref{fig:impl}. Next, the SP configures the PPC to define non-secure access settings for each NSVW's devices, and, lastly, the SP configures the NVIC's ITNS registers to re-direct the interrupts of the first NSVW to the normal world. 

\mypara{\textcircled{3} {uTango} Kicking-off.} \label{ui6}
In the last boot stage, \textsc{uTango} is responsible for configuring the CPU state for the first NSVW and kick off the execution. A non-secure call is issued to the entry point of the NSVW \#1 (Figure \ref{fig:lifecycle}). This function will switch the CPU state from secure to non-secure by issuing a BLXNS instruction. All the register banks are cleared to avoid information leakage. The boot sequence is complete at this stage, and the processor starts executing the first NSVW. 


\mypara{\textcircled{4} WS Scheduling.} \label{ui7}
At \textit{run-time}, \textsc{uTango} is mainly responsible for scheduling and context switching NSVWs. The WS keeps the suspended NSVW states and resources in the secure world while remapping the next-to-run NSVW resources as non-secure. The WS process consists of four main steps (4.1-4.4). 
In the first step (4.1), the WS saves the processor context of the suspended NSVW. Thus, all general-purpose and special CPU registers, as well as selective SCB registers, are stored in the respective WCB. The WS is implemented in assembly, enabling these multiple accesses to the WCB memory segment to be combined into a single store-multiple instruction (STM) to improve performance. The NVIC state is also preserved in the WCB interrupt descriptor (details in Section \ref{ui8}). Next (4.2), the scheduler selects the next-to-run NSVW, according to a round-robin policy. The WS retrieves the stored SAU data table from the scheduled NSVW's WCB entry to program SAU in step three (4.3). The SAU re-configuration also leverages fine-grain assembly customizations by leveraging load-multiple instructions (LDM). However, due to the lack of fast-reconfiguration optimization mechanisms available in the SAU, the WS needs to program all eight SAU regions by iteratively accessing the RNR register. Finally, in step (4.4), the context of the scheduled NSVW is loaded to the CPU. At this point, a branch is issued, and processor execution is switched to the non-secure state. 

\subsection{Worlds Interrupt Handling}\label{ui8}

In TrustZone-M MCUs, the NVIC registers are not banked between security states. The ITNS register enables the configuration of the interrupt's security target. Once an interrupt is configured as secure, accesses to the associated fields in non-secure aliases are read-as-zero. Thus, NVIC's non-secure state must be preserved for each interrupt assigned to an NSVW. The WCB structure features a descriptor that holds the Interrupt Set Enable Register (ISER), Interrupt Set Pending Register (ISPR), Interrupt Priority Register (IPR), and ITNS. Interrupt management was first implemented using a non-preemptive mechanism where interrupts are served as soon as the respective NSVW is scheduled. In this case, i.e., the worst-case scenario, the interrupt latency is delayed by the amount of time needed to perform a complete round of NSVWs (i.e., \mbox{$((NSVW-1) * tick)+schedtime$}). However, for real-time applications, this latency may be prohibitive. 
Current efforts are going through the extension of \textsc{uTango} to implement a preemptive priority-based mechanism. In the following, we only explain the current implementation, but in Section \ref{dis}, we explain the challenges of implementing the preemptive mechanism. However, for the sake of clarity, we illustrate an example of the execution flow of both approaches in Figure 4.

\begin{figure}[t]
    \centering
    \begin{subfigure}[b]{0.47\textwidth}        
        \centering
        \includegraphics[width=1\linewidth]{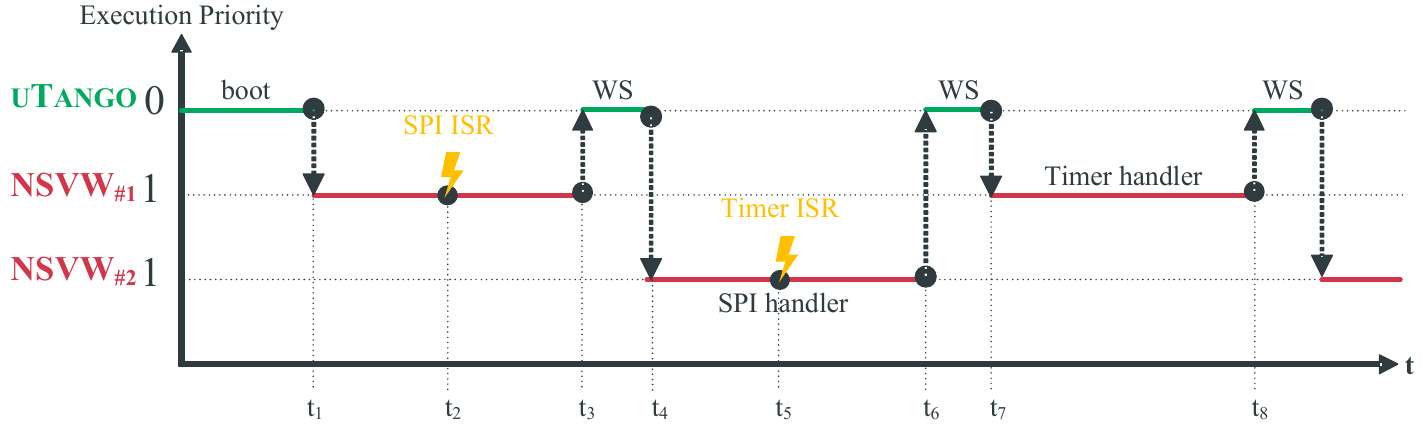}
        \caption{Current non-preemptive world interrupt handling.}
        \label{fig:int1}
    \end{subfigure}
    \begin{subfigure}[b]{0.47\textwidth}        
        \centering
        \includegraphics[width=1\linewidth]{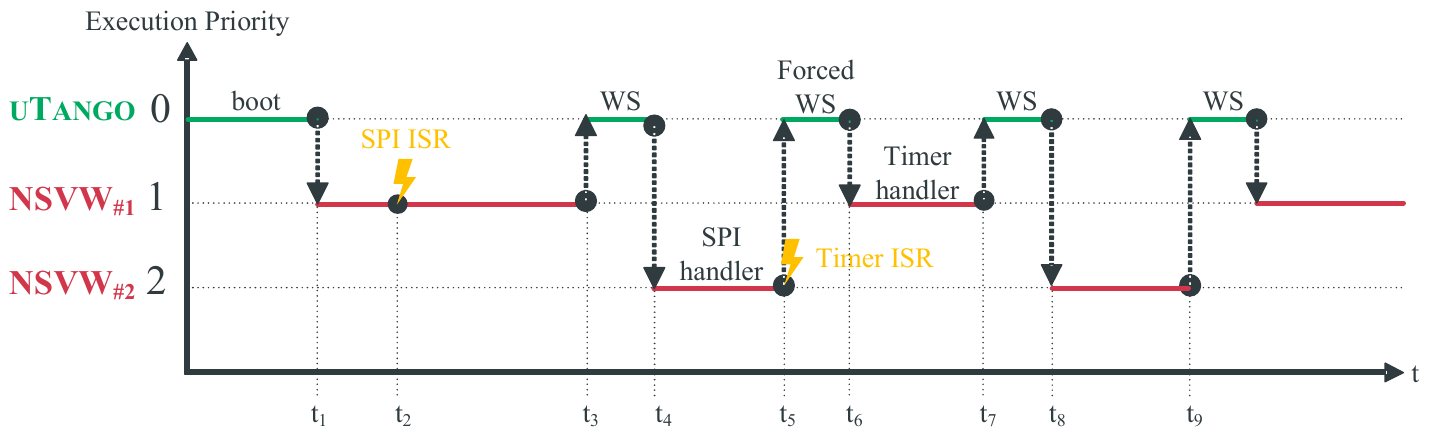}
        \caption{Near-future priority-based world interrupt handling.}
        \label{fig:int2}
    \end{subfigure}
    \caption{\textsc{uTango} world interrupt handling mechanisms.}
    \label{fig:int}
\end{figure}

\mypara{Non-preemptive World Interrupt Handling.} \label{ui9}
Figure \ref{fig:int1} illustrates the non-preemptive interrupt handling flow. The vertical axis depicts the execution environment and its respective priority. \textsc{uTango} is represented as the system's higher-priority workload (smaller number in priority level) since, by design, the secure world is more privileged than the normal world. NSVW \#1 and NSVW \#2 have the same priority level and take equal CPU quantum. A timer interrupt is assigned to NSVW \#1, while an SPI interrupt is assigned to the NSVW \#2. As previously explained, interrupts assigned to non-executing NSVWs, such as the SPI triggered at $t_2$, will only be served as soon as its respective NSVW is put into context (from $t_4$ to $t_6$). For this particular case, the SPI interrupt latency is equal to $t_4 - t_2$. The same behavior happens to the timer interrupt. The interrupt is triggered at $t_5$ (while NSVW \#2 is running) but just served when NSVW \#1 is resumed at $t_7$. \textsc{uTango} enables this non-preemptive behavior by saving and restoring each interrupt state during the context switching of NSVWs ($t_3$ to $t_4$ and $t_6$ to $t_7$). In particular, at the first scheduling point ($t_3$ to $t_4$), the WS will save the NVIC's state for the timer interrupt and restore the SPI's pending bit and security target before resuming the NSVW \#2. Once NSVW \#2 is resumed at $t_4$, if the SPI interrupt is enabled, the pended request will force the processor to attend the interrupt and start executing the respective SPI's interrupt handler. 

Figure 4b illustrates the priority-based interrupt handling flow. Each NSVW is assigned (i.e., system designer) with a priority according to its criticality level. Higher-priority NSVWs will preempt the processor and grant execution to attend asynchronous events. In contrast, low-priority NSVWs will be blocked from interrupting the processor during the execution of high-priority workloads.


\subsection{Worlds Communication}

\textsc{uTango} implements a secure blocking and non-blocking request-response messaging mechanism (read/write APIs for each protocol - 4 in total) to enable NSVWs to communicate with each other. A blocking operation forces WCC to schedule the callee-NSVW, and the API will only return once the NSVW responds, thus completing the operation. Alternatively, when a caller-NSVW issues a non-blocking operation, the callee-NSVW only responds when it gets scheduled by the WS. The messages exchanged in these channels are limited to a 12-byte data stream and are sent via registers. The WCC uses the internal inbox structure of each NSVW to pass messages across senders and receivers. These APIs are implemented through secure entry points located in a pre-defined NSC memory region (light-blue NSC region in Figure \ref{fig:impl}), i.e., the WCC gateway. When an NSVW uses the sending API, the 12-byte data stream is copied to registers (i.e., \textit{r4-r6}), which are read and placed into the receiver's inbox by the WCC. When the receiver NSVW reads the message, it calls the WCC via the receiving API that copies the inbox message (if full) to the registers. The WCC carefully avoids information leakage by clearing the remaining CPU registers before returning to the active NSVW. 



\subsection{Reference IoT Application}
\begin{figure}[t]        
    \centering
    \includegraphics[width=1\linewidth]{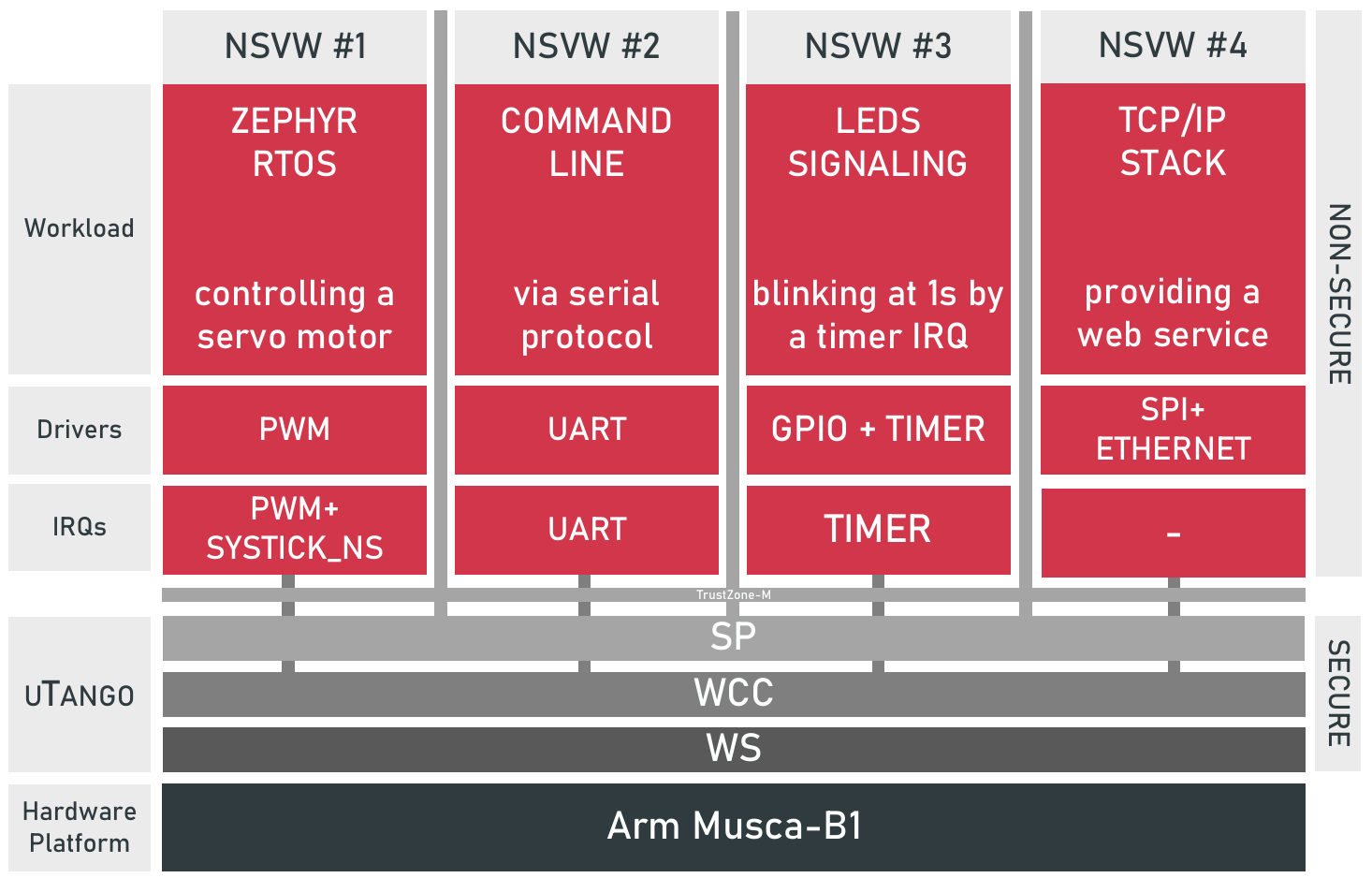}
    \caption{\textsc{uTango} IoT reference application.}
    \label{fig:refapp}
\end{figure}

Figure \ref{fig:refapp} depicts the \textsc{uTango} IoT reference application. Typically, IoT-based applications are monolithic firmwares composed of well-defined and easily isolable 3rd party libraries that are linked and compiled into a bloated binary. \textsc{uTango}'s unique design allows system developers to separate each single building block into an isolated environment. Therefore, from a system developer perspective, they must to first (i) identify the application's sub-components (and their resources) that need to be functionally protected, (ii) integrate the building block in an isolated NSVW that offers a stripped \textit{main} application, (iii) define each resource property of the target NSVW in the configuration file (memory regions, devices, interrupts, and communication channels) and (iii) finally, use the inter-NSVW communication channels (i.e., the blocking or non-blocking APIs) to exchange messages between NSVWs. 

\par In our reference application, we selected a set of building blocks aiming at demonstrating the applicability of \textsc{uTango} to develop secure IoT devices. These building blocks implement the main features required by IoT devices, ranging from secure connectivity, real-time operation, and local management. Specifically, in Figure \ref{fig:refapp}, the proof of concept system controls - via a local terminal console (NSVW \#2) - a servo motor, operating under an RTOS (i.e., Zephyr in NSVW \#1). The RTOS receives commands from the local terminal console via the secure non-blocking APIs provided by \textsc{uTango's} WCC. The NSVW \#4 implements a full-blown TCP/IP stack that can receive commands from the other NSVWs to send data to a web service. The NSVW \#3 blinks a LED at each timer overflow interrupt to demonstrate the world's interrupt handling mechanism. The \textsc{uTango}'s SP provides hardware-enforced separation among all consolidated workloads.

\section{Security Analysis}
\label{sa}

\par TrustZone-M hardware primitives ensure hardware-enforced isolation of system resources, i.e., code, data, devices, and interrupts. Thus, TrustZone-M hardware hooks are sufficient to materialize the \textsc{uTango}'s vision of providing a new security model that allows the consolidation of multiple, equally-secure, execution environments. Despite offering a strong foundation for system-level security, such as data and code protection, TrustZone lacks in defining anti-tampering mechanisms and side-channel protection. Therefore, physical and side-channel attacks are out-of-the-scope of this work. Hence, we trust Arm to provide a non-compromised TrustZone-M design in the Musca-B1 platform. 

\par Our threat model is based on the very same assumptions delivered by the TrustZone-M \cite{Arm2018, ArmKeil2019}. 
Firstly, we assume that an adversary cannot escalate an attack to the secure software (i.e., \textsc{uTango}) through a compromised piece of non-secure software (i.e., an NSVW) via a local or remote software attack. Secondly, our multi-world security model takes advantage of the same hardware mechanisms that isolate the secure and non-secure worlds to guarantee that a compromised NSVW cannot hijack other execution environments. In a nutshell, we must assume the following:

\begin{itemize}
    \item NSVWs running on the non-secure side are untrusted.
    \item \textsc{uTango} kernel, including all its software components, e.g., WS, and SP, are trusted.
    \item TrustZone-M hardware security extensions guarantee strong isolation between secure and non-secure states.
    \item The misuse of the TrustZone-M hardware components, such as the SAU and IDAU, can compromise the security model. Thus, the configuration file which defines the overall system partition must be trusted. 
\end{itemize}

\par Next, we discuss how \textsc{uTango} can mitigate potential attack vectors that a malicious adversary could exploit. 

\mypara{Protection of {uTango} kernel.} \label{ev-sa2}
Enabled by TrustZone hardware controllers, the first isolation layer of the system separates the \textsc{uTango} kernel from the NSVWs. When the kernel partitions the system using the SAU, it ensures that all NSVWs run in the normal world, while the kernel is kept protected in the secure world. Therefore, the SAU controller validates all NSVW's memory transactions, blocking access to resources outside of its domain. NSVWs can also attempt to extract information from \textsc{uTango} by leaking non-banked CPU registers. The attack can be carried out after a state transition, i.e., from secure to non-secure, which can happen after the (i) boot stage or (ii) after context switching periods. The kernel prevents the leakage by (i) clearing all registers before jumping to the non-secure state and by (ii) inherently replacing the CPU state with the scheduled NSVW context. 

\mypara{Isolation of NSVWs.} \label{ev-sa3}
The \textsc{uTango} partition prevents NSVWs to share memory regions, devices, and interrupt sources. The bi-directional isolation of the NSVWs is enforced by the SAU and additional SoC security gates. In TrustZone-M platforms the configuration of the SAU and security gates is restricted to the secure world and managed by the \textsc{uTango} kernel. The \textsc{uTango} architecture allows developers to easily deploy 3rd party workloads. This feature can be leveraged by attackers to install malicious libraries or applications. Nevertheless, \textsc{uTango} hardware-enforced isolation prevents a compromised NSVW from overcoming its boundaries and escalating to other domains. 

\mypara{Secure Bus Masters.} \label{ev-sa5}
As discussed in Section \ref{ui2}, we assume that NSVWs interactions with additional bus masters are protected and mediated by the \textsc{uTango}. This design decision avoids the overhead of configuring each bus filter (i.e., MPC, PPC) during context switching, saving a considerable amount of CPU clock cycles. Moreover, this also prevents an attacker from gaining access to the overall system's secure or non-secure memory. However, in an application scenario where an NSVW needs access to a DMA controller, \textsc{uTango} will offer a set of secure services to interface such modules.

\section{Evaluation}
\label{ev}
We evaluated \textsc{uTango} on an Arm Musca-B1 Test Chip Board, which features two Cortex-M33 processors, running at 40 MHz (CPU0) and 160 MHz (CPU1). Firstly, in Section \ref{ev_ws}, we assess a set of security metrics derived from the BenchIoT suite \cite{Bagchi2019}. On Section \ref{ev3}, we evaluate performance, and in Section \ref{ev4} we focus on the interrupt latency. Lastly, in Section \ref{ev5}, we evaluate code and binary sizes.

\subsection{Security Metrics} \label{ev_ws}
\par BenchIoT \cite{Bagchi2019} is a recent benchmark suite and an evaluation framework to evaluate security solutions for IoT-based MCUs. The suite enables the automatic collection of 14 metrics for security, performance, memory usage, and energy consumption. As of this writing, BenchIoT can only support Armv7-M architectures \cite{Bagchi2019}, i.e., BenchIoT cannot be used to evaluate \textsc{uTango}. Notwithstanding, we performed a best-effort evaluation of \textsc{uTango}'s security based on the BenchIoT's eight security metrics while keeping as close as possible to the framework principles and metrics criteria. 
\par According to the evaluation model presented in Ref. \cite{Bagchi2019}, we organized the security metrics in three goals: (i) minimizing privileged execution (i.e., a total of privileged and system call cycles); (ii) enforcing memory isolation (i.e., maximum code and data region ratio); and (iii) control-flow hijacking protection (i.e., number of available ROP gadgets and indirect calls, and data execution prevention).

\mypara{Minimizing Privileged Execution.}
For Armv7-M MCUs, BenchIoT counts as privileged cycles all instructions executed in privileged thread and handler mode. For TrustZone-enable Armv8-M MCUs, all CPU modes are banked, but the secure world is always considered more privileged than the normal world. Thus, we count all instructions executed in the secure privileged thread and secure handler mode as privileged cycles. In the context of our architecture, \textsc{uTango} is the single component running in secure privilege thread mode, at boot time, and secure handler mode, at run-time. All NSVWs run within the normal world, so all execution cycles are not considered.
Our results show that, at run-time, \textsc{uTango} runs a total of 215 privileged cycles at each system tick (Section \ref{ev31}). At boot-time, the total number of thread privileged cycles is 7749 cycles (for 1 NSVW) and increases, on average, 1236 cycles for each extra added NSVW.
BenchIoT also counts the number of SuperVisor Call (SVC) cycles. Unprivileged code can leverage this instruction to trigger a system call intended at executing privileged thread code. This mechanism can be leveraged as a potential attack vector. In the context of TrustZone-M MCUs, SVC calls can be issued in secure and non-secure states. As aforementioned, non-secure SVCs are not considered as normal world code is always considered non-privileged. \textsc{uTango} does not issue any SVC call, i.e., the number of secure SVC calls is zero. The execution flow has a well-defined entry and exit point, always running in secure handler mode. 


%

\mypara{Enforcing Memory Isolation.}
Another two security metrics evaluated by BenchIoT are the (i) maximum data region ratio and (ii) maximum code region ratio. These two metrics aim at assessing memory isolation's effectiveness by computing the size ratio of the maximum available code/data regions to an attacker with respect to the total code/data size of the application binary \cite{Bagchi2019}. \textsc{uTango} isolates each environment within a strong compartment, with boundaries of the NSVW defined per binary needs. Thus, the maximum data and code region ratio is 0. 

\mypara{Control-Flow Hijacking Protection.}
Code reuse attacks (i.e., ROP gadgets and indirect calls) are among the most common attack vectors used to hijack the control flow of an application. To measure the number of ROP gadgets, we used the ROPGadget tool. \textsc{uTango} has a total of 303 ROP gadgets. Notwithstanding, a deeper investigation unveiled that these ROPs belong to the boot-related code and are not executed during runtime. As stated in Section \ref{ui3}, all the world scheduling logic is implemented in assembly. Although this number is an order of magnitude smaller compared to ATF-M and the results presented in Ref. \cite{Bagchi2019}, we are aiming at squeezing this value in the near future by also implementing the boot logic in assembly or leveraging inline substitution optimizations. Indirect calls are another type of code reuse attacks that relies on using function pointers to hijack the control flow. In the case of \textsc{uTango}, we parsed the binary file, and we found only one indirect call related to the secure to non-secure exit point, issued through a BLXNS instruction. 

Another important aspect in defending against control-flow hijacking is related to data execution prevention (DEP) mechanisms. In the context of Arm MCUs, proposed defense mechanisms leverage the MPU to enforce memory regions, either writable (data) or executable (code) \cite{Bagchi2019}. \textsc{uTango} currently leverages the secure MPU to enforce DEP among kernel code and data sections. Furthermore, as mentioned in Section \ref{ev-sa7}, we will also leverage the secure MPU to enforce isolation and deploy a DEP defense mechanism among security gates.

\begin{table}[t]
\centering
\scalebox{0.70}{
\begin{tabular}{ p{1.7cm} | p{2.1cm} || p{1.2cm} | p{4.1cm}}

\hline
\multicolumn{2}{c ||}{\cellcolor{lightgray!50}\textbf{Platform}} &  \multicolumn{2}{c}{\cellcolor{lightgray!50}\textbf{Toolchain}}  \\
\hline

\multirow{2}{*}{frequency}       &   \multirow{2}{*}{\small{40 MHz (CPU0)}}       &   Compiler version            &  \small{GNU Arm Embedded Toolchain arm-none-eabi-gcc 9.3.1} \\ 
\hline
max. frequency             &   \multirow{2}{*}{\small{160 MHz (CPU1)}}      &   Linker version              & 	\small{GNU binutils ld version 2.34.0}  \\
\hline
architecture                &   \small{Armv8-M}             &   \multicolumn{2}{c}{\cellcolor{lightgray!50}\textbf{Flags}} \\
\hline
isa                         &   \small{Thumb/Thumb-2}       & \multirow{3}{*}{compiler}     &  \multirow{3}{4.1cm}{\small\textit{-Os -march=armv8-m.main -mcpu=cortex-m33+nodsp -ffunction-sections -mfloat-abi=softfp -mthumb}}   \\
\cline{1-2}
address size                &	\small{32-bit}              &                               &  \\
\cline{1-2}
code memory size         & \multirow{3}{*}{\small{512 KiB eSRAM}}           &                               & \\
\hline
data memory size            & \multirow{2}{*}{\small{512 KiB iSRAM}}           & \multirow{3}{*}{linker}       & \multirow{3}{4.1cm}{\small\textit{-O2 -Wl,-gc-sections -march=armv8-m.main -mcpu=cortex-m33+nodsp -mfloat-abi=softfp -mthumb -specs=nosys.specs}}\\
\cline{1-2}
processor name              &   \multirow{2}{*}{\small{Cortex-M33}}          &                               &   \\
\cline{1-2}
caches                      &   \small{2 KiB ICache/core}             &                               &   \\
\hline
 active cores               &   \small{1 (CPU0)}                   &  libs                         &   \small\textit{{'user libs': ['-lm']}} \\ 
\hline

\end{tabular}
}
\caption{Platform, toolchain, and compilation details.}
\label{tb-embench}
\end{table}


\subsection{Performance Overhead}
\label{ev3}

\begin{figure*}[!ht]   
    \begin{subfigure}[b]{1\linewidth }        
        \centering
        \includegraphics[width=1\linewidth]{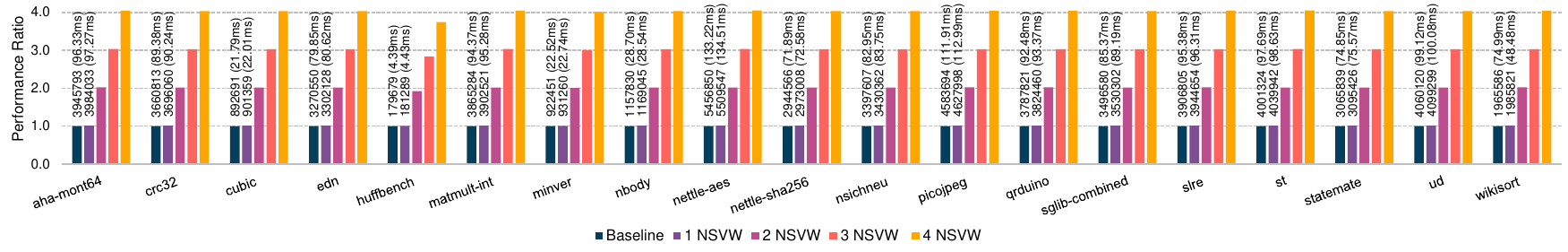}
        \caption{0.5 ms \textsc{uTango} tick.}
        \label{fig:embench1}
    \end{subfigure}
    \begin{subfigure}[b]{1\linewidth }        
        \centering
        \includegraphics[width=1\linewidth]{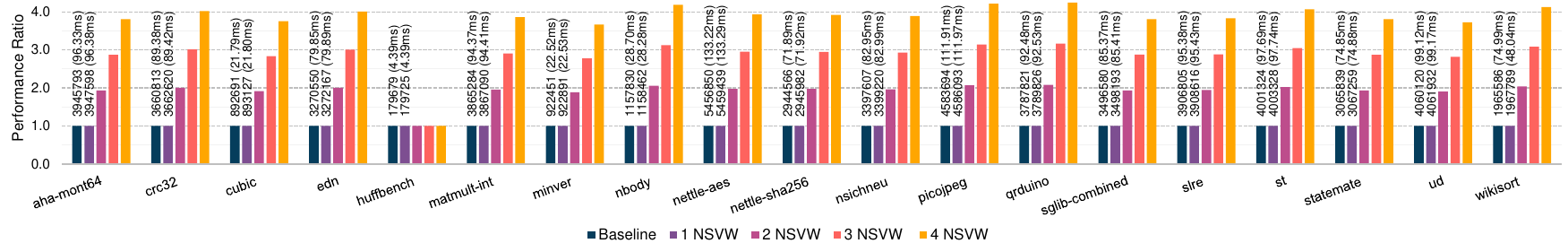}
        \caption{10 ms \textsc{uTango} tick.}
        \label{fig:embench2}
    \end{subfigure}
    \caption{Performance overheads (ratio) of Embench benchmark suite relative to bare-metal execution.}
    \label{fig:embench}
\end{figure*}

\subsubsection{World Switch Time}\hfill \label{ev31}

\mypara{Experimental Setup.}
The world switch time is defined as the amount of time that the \textsc{uTango} kernel takes to switch between NSVWs. As explained in Section \ref{ui7}, this operation includes saving and restoring the worlds' context (i.e., core registers, system registers, and NVIC), re-configuring the SAU regions (to enforce memory isolation), and running the scheduler algorithm. We used the Data Watchpoint and Trace (DWT) unit from the CoreSight debug system to measure the world switch time, which features a 32-bit cycle counter running at the CPU clock frequency. The DWT cycle counter is read before and after completing the WS operation.

\mypara{Results and Conclusions.}
During our measurements, we collected 1000 samples, and all the collected samples reported a world switch time of exactly 215 clock cycles, i.e., 5.4 microseconds (µs) at 40 MHz. 
The high determinism is a reflex of (i) the characteristics of the Armv8-M architecture and (ii) from the fact \textsc{uTango} runs from a TCM (Section \ref{ui4}). The reduced world switch time is a consequence of (i) the raw assembly implementation of the WS logic and the (ii) TCM. For instance, when configured with a 10 milliseconds (ms) tick rate, the expected performance penalty is a negligible 0.054\%. These results (i) are from the same order of magnitude of MultiZone commercial solution (i.e., MultiZone's zones switch penalty is 175 clock cycles) and (ii) some specific use cases may minimize transitions between worlds (e.g., wfi/sleep modes).

 


\subsubsection{Run-time Overhead}\hfill \label{ev32} 

\mypara{Benchmark Suite.}
To evaluate the run-time overhead, we used Embench (version 0.5) \cite{embench20}. Embench is a free and open-source benchmark suite specially designed for deeply embedded systems. Assuming the presence of no OS and minimal C library support, Embench targets small devices with a few kilobytes of Flash (ROM) and RAM. Embench consists of 19 real programs, representatives of the following metrics: branch, memory, and computing requirements.
Each benchmark reports a single summarizing performance score that outputs the geometric mean and geometric standard deviation ratios relative to a reference platform or setup, which in our case represents the Musca-B1 Test Chip Board. 

\mypara{Experimental Setup.} 
Despite Arm Musca-B1 featuring a dual asymmetric Cortex-M33 MCU, \textsc{uTango} currently only supports a single-core configuration. Thus, in our experiments, we have only enabled the CPU0, running at 40 MHz.
We ran the benchmarks natively on the target platform. Then, each benchmark was executed with \textsc{uTango}, configured to support 1, 2, 3, and 4 NSVWs, which represents a reasonable number of environments on a typical application scenario.
We compiled each benchmark to the target platform and ran them unmodified in the first NSVW environment with the configuration described in Table \ref{tb-embench}.
%
The other NSVWs were running a toy bare-metal application, implementing a bare infinite loop. These worlds do not yield (i.e., they consume all their CPU quantum) and, therefore, even if a more realistic application is used, the benchmark results would be the same.
The DWT cycle counter is used to measure the total number of clock cycles taken to complete the benchmark (i.e., trigger points are placed before the beginning of the benchmark operation and after it finishes).
Experiments were repeated for different tick rate configurations, ranging from 0.5 ms to 10 ms. The achieved results for 0.5 and 10 ms are illustrated in Figure \ref{fig:embench}, where each bar (representing 1, 2, 3, and 4 NSVWs) depicts the respective ratios relative to the baseline. On top of the first and second bar, the absolute execution time, in clock cycles, and total execution time in ms (within parentheses) is also presented. 

\mypara{Results and Conclusions.}
Looking at Figure \ref{fig:embench}, we can draw four main conclusions, discussed throughout the following paragraphs. 

\mysubpara{1}{\textit{Residual overhead with 1 NSVW:}}
\noindent With a 1 NSVW configuration, the overhead introduced by \textsc{uTango} is almost residual. For instance, for a 10 ms tick rate, the average performance overhead is 0.05\%, which is within the expected theoretical overhead (see world switch time). For a 500 µs tick rate, which is considered an unusual high switching rate (i.e., highly responsive system), the average performance overhead is less than 1\%. 

\mysubpara{2}{\textit{Linearly overhead with the increase of NSVWs:}}
\noindent From the conducted experiments, we can observe that the performance overhead increases (almost) linearly with the number of NSVW's, i.e., the third, fourth, and fifth bars (2, 3, and 4 NSVWs, respectively) increase the performance overhead by a similar ratio. 
We observed this same phenomenon in other testbed scenarios with more than 4 NSVWs, which we have decided not to present due to space limitations. This impact is expected and is a natural consequence of sharing the CPU among all NSVWs, i.e., each NSVW gets a CPU quantum equal to the tick rate.

\mysubpara{3}{\textit{Relative overhead decrease with higher tick rates:}}
A third and interesting observation is related to the heterogeneity in the performance overhead ratio when the system is configured with multiple NSVWs, particularly for 3 and 4 NSVWs. This phenomenon becomes even more evident for the experiments conducted with a 10 ms tick rate. We observed that increasing the \textsc{uTango} tick rate may suggest that, for the majority of the benchmarks, the system gets an increase of performance (i.e., a decrease of the performance overhead ratio). The most evident case is observed for the \textit{huffbench} benchmark, which for a 10 ms tick rate suggests that there is no performance penalty, i.e., the ratio is always 1 no matter how many NSVWs are running in the system. While this may look a bit surprising at first sight, a deeper investigation pointed out that this is a consequence of the execution time of the benchmark. As presented on top of the first two bars (in parentheses), the native execution time of benchmarks ranges from a few ms to dozens of ms. For instance, the bare execution of the \textit{huffbench} takes around 4.39 ms while the \textit{nettle-aes} takes 133.22 ms. So, when the system shares the CPU among multiple NSVWs, and depending on the tick rate, the benchmark may finish in fewer rounds, decreasing the performance overhead ratio. We repeated the experiments also for a tick rate of 1, 2, and 5 ms, and these again show very clearly the explained pattern. 

\mysubpara{4}{\textit{Overhead increase in corner cases:}}
Finally, and as complementary to the phenomenon described above, we also observe that there are some exotic benchmarks, i.e., \textit{crc32}, \textit{nbody}, \textit{picojpeg}, and \textit{qrduino}, that present an increase of the performance overhead ratio. In this case, this apparent increase of performance overhead is justified by the warm-up time of the benchmark. Thus, when a higher warm-up time is required, most of the available time slots in the first round are wasted, which results in an additional impact on the final performance overhead ratio. 

%

\begin{figure}[t]
    \begin{subfigure}[b]{1\columnwidth }        
        \centering
        \includegraphics[width=1\columnwidth]{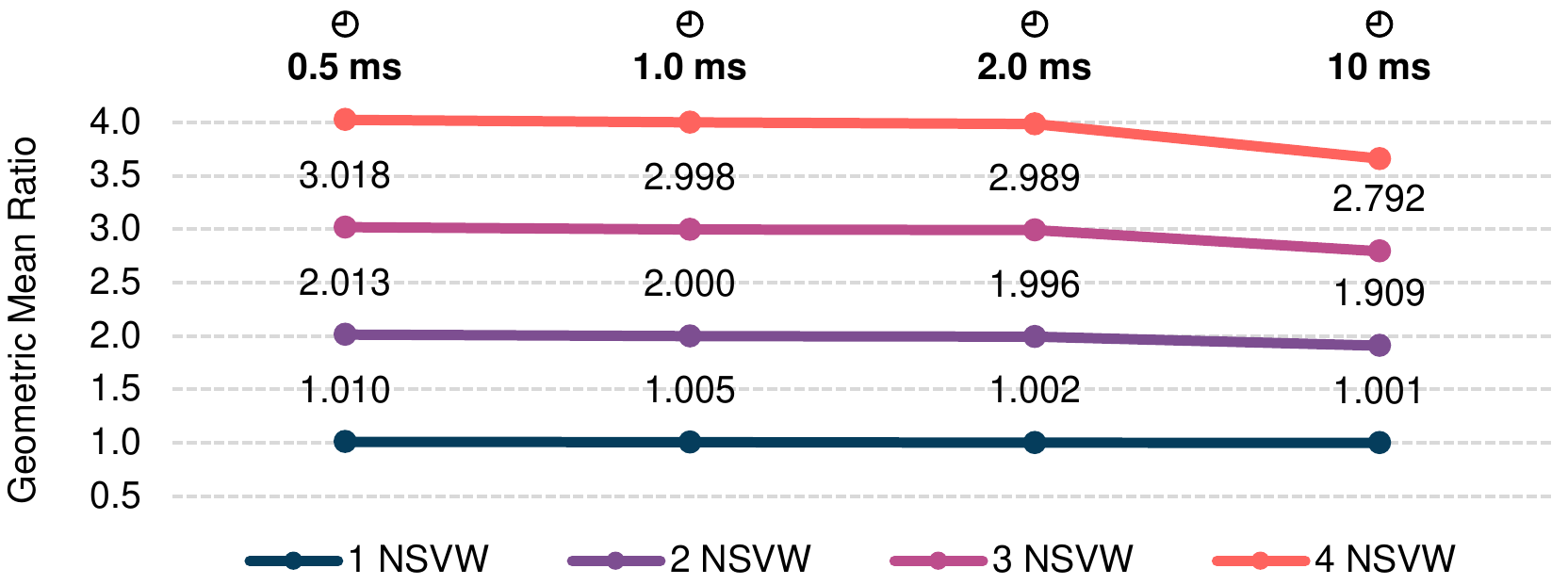}
        \caption{Performance overhead for 1, 2, 3, and 4 NSVWs.}
        \label{fig:tick234}
    \end{subfigure}
    \begin{subfigure}[b]{1\columnwidth }        
        \centering
        \includegraphics[width=1\columnwidth]{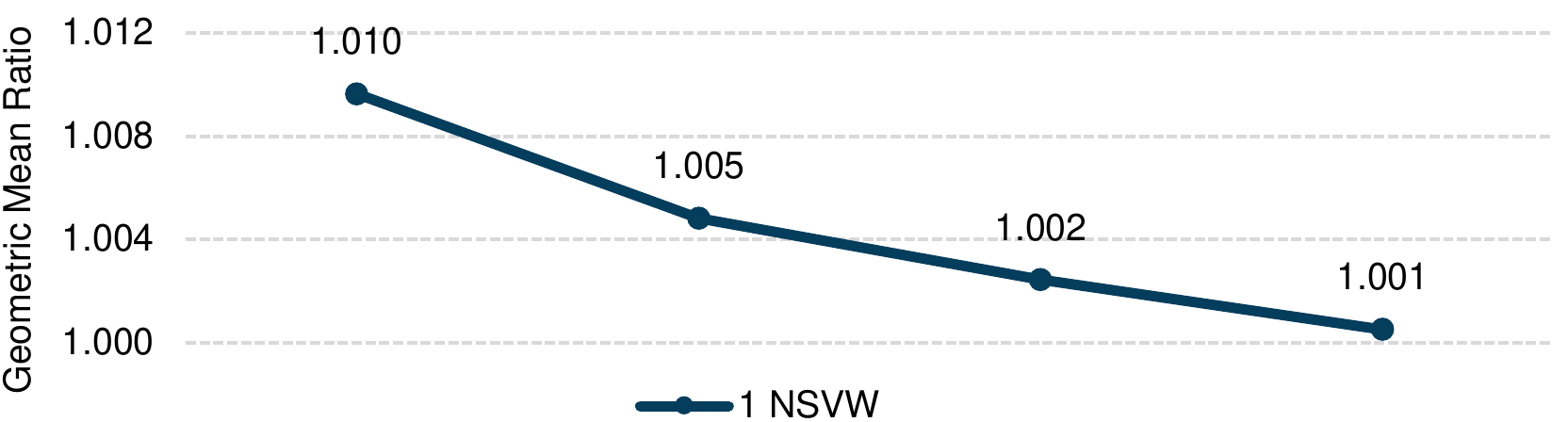}
        \caption{Zoomed view of the performance overhead for 1 NSVW.}
        \label{fig:tick1}
    \end{subfigure}
    \caption{Performance overhead vs variation of \textsc{uTango} tick for different configurations.}
    \label{fig:tick}
\end{figure}

\mypara{Exploring further C3.}
Figure \ref{fig:tick} depicts the impact of the \textsc{uTango} tick rate variation on the overall performance overhead. 
We have repeated this experiment for four different tick rates (0.5 ms, 1 ms, 2 ms, and 10 ms), where each obtained value corresponds to the geometric mean ratio for the full-run of the Embench suite running in systems configured with different NSVWs. 
From the obtained results, we can validate the phenomenon described in \textbf{C2}. While increasing the \textsc{uTango} tick rate, there is an apparent decrease of the performance overhead (Figure \ref{fig:tick}). Looking at Figure \ref{fig:tick1}, we can also conclude that the performance overhead increases exponentially while decreasing the tick rate. However, this exponentially increase is highly acceptable because, for a 500 µs tick rate, the performance overhead is less than 1\%. This impact will be less noticeable in platforms running at higher frequencies, e.g., the NXP LPC55S69-EVK and STM NUCLEO-L552ZE-Q, which run approximately three times higher, and, therefore, the overhead would decrease by a factor of three. 

\begin{figure}[t]
    \centering
    \begin{subfigure}[b]{\linewidth }        
        \centering
        \includegraphics[width=1\linewidth]{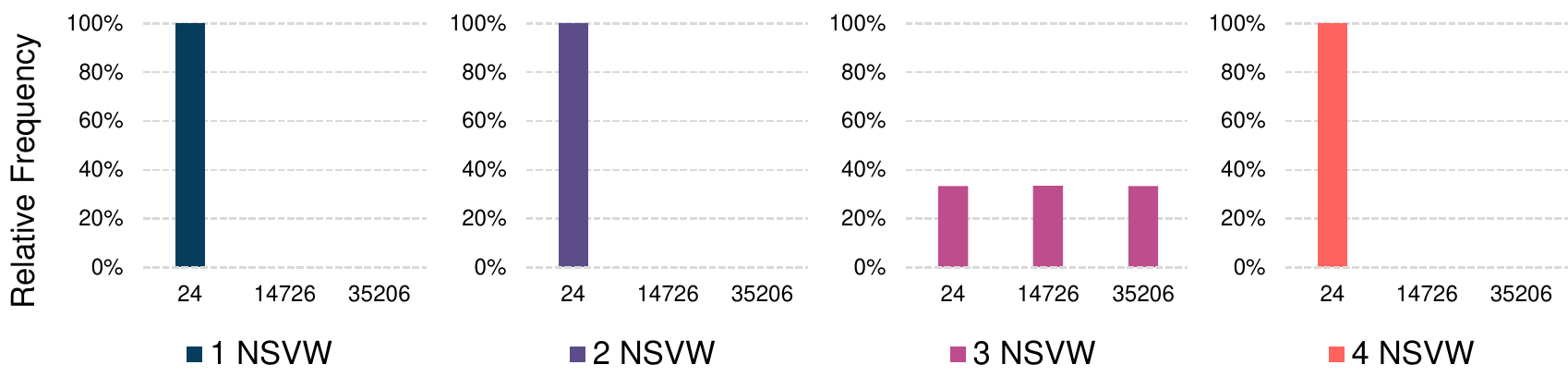}
        \caption{\textsc{uTango} with 0.5 ms tick.}
        \label{fig:int_lat1}
    \end{subfigure}
    \begin{subfigure}[b]{\linewidth }        
        \centering
        \includegraphics[width=1\linewidth]{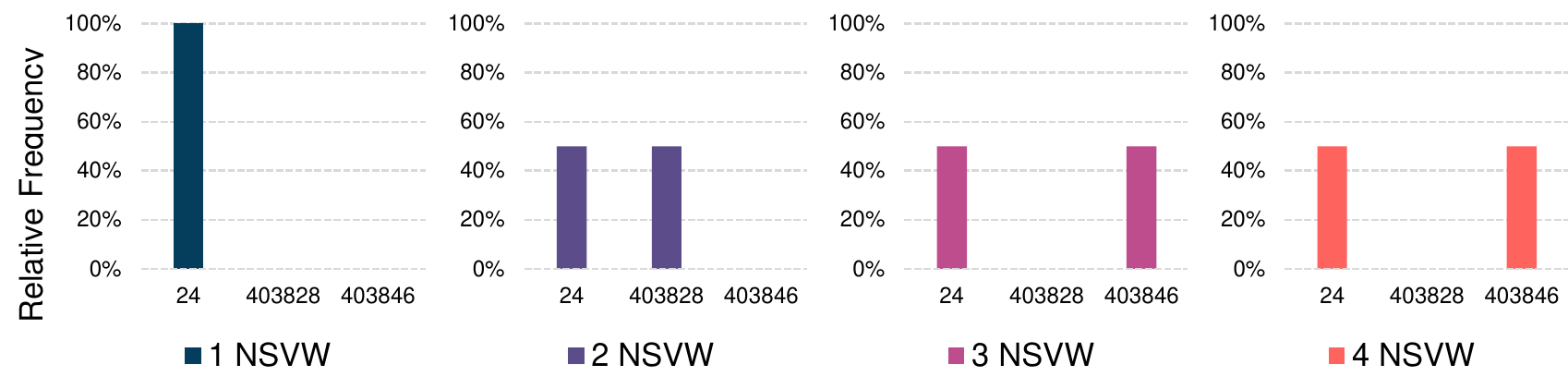}
        \caption{\textsc{uTango} with 10 ms tick.}
        \label{fig:int_lat2}
    \end{subfigure}
    \caption{Relative frequency of interrupt latency expressed in clock cycles.}
    \label{fig:int_lat}
\end{figure}

\subsection{Interrupt Latency}
\label{ev4}

To measure the interrupt latency, we crafted a minimal bare-metal benchmark application, running in the NSVW\#1, that continuously configures a timer (25 nanoseconds resolution) to trigger an interrupt every 10 ms. Since the time stamp when the interrupt was triggered is known, the latency can be calculated as the difference between the expected wall-clock time and the actual instant it starts handling the interrupt. 
We have performed two experiments for different \textsc{uTango} tick rates (0.5 ms and 10 ms) while varying the number of NSVWs to be scheduled. The results were obtained by taking 1000 samples.
Figure \ref{fig:int_lat} depicts the relative frequency of each interrupt latency measurements.  
The results are expressed in the number of clock cycles required by the CPU to start executing the timer handler. 
According to Figure \ref{fig:int_lat}, we can drawn two major conclusions.

Firstly, depending on when the interrupt is triggered, we can achieve better or worse execution times: (1) if the interrupt is triggered while the NSVW handler is executing, the measured latency is near its native values (observed for all configurations with one world). The NSVW receives the interrupt transparently through normal hardware interrupt behavior, and the final interrupt handler executes within a fixed 24 clock cycles; 
(2) if the interrupt triggers when a different NSVW is active, the interrupt latency increases significantly since the interrupt will only be handled when the assigned NSVW is scheduled.

The second takeaway comes from observing a direct relation between the system tick rate, the number of NSVWs, and the time interval selected for the interrupting timer (10 ms). 
Considering the scenario with 2 NSVW, depicted in Figure \ref{fig:int_lat1}, the interrupt latency for the 1000 collected samples shows a relative frequency of 100\% for 24 clock cycles. 
Such results are explained by the number of rounds (each one taking 1 ms) needed to complete the 10 ms interval (at a tick rate of 0.5 ms) when two worlds are configured, which is 20. 
This means that the timer interrupt will be triggered approximately when the handling world is executing. 
On the other hand, in the scenario with 3 NSVW, each round takes 1.5 ms to complete; therefore, the system needs $\approx$6.67 rounds to complete 10 ms. 
Such variation shifts the trigger point of the interrupt to a specific point in time where a different NSVW is executing, delaying the interrupt to be serviced. 



\subsection{Code and Binary Size}
\label{ev5}

\begin{table}[t]
\centering
\scalebox{0.76}{
\begin{tabular}{l||lll||llll}
\hline
\textbf{directory}         & \multicolumn{3}{c||}{\textbf{SLoC}} & \multicolumn{4}{c}{\textbf{size (bytes)}} \\
                           & C       & asm    & total           & .text   & .data   & .bss  & total          \\ \hline
 arch/armv8-m               & 787     & 393    & 1180            & 1336    & 0       & 0   & 1336           \\
 platform/MUSCAB1           & 1140    & 0      & 1140            & 1048    & 604     & 0     & 1652           \\
 core                       & 264     & 0      & 264             & 552    & 0       & 652   & 1204           \\
 config \textit{(2 worlds)} & 93     & 0      & 93             & 0       & 108     & 0     & 108            \\
 \hline
 total                      & 2284    & 393    & \textbf{2677}   & 2936    & 712     & 652   & \textbf{4300}  \\ \hline
 \end{tabular}
 }
 \caption{Source lines of code (SLoC) and binary size (bytes) by directory.}
 \label{tb-loc}
 \end{table}

\par \textsc{uTango} was developed from-scratch with no dependencies on compiler or external libraries. Table \ref{tb-loc} reports (i) the number of SLoC and (ii) the binary size. 

\mypara{Source lines of code.} 
To count the number of SLoC, we used the SLOCCount tool. \textsc{uTango} implementation code is divided into three main directories: (i) \textit{\textbf{arch}}, targetting Armv8-M architectural-specific functionalities; (ii) \textit{\textbf{platform}}, containing platform-specific code (e.g., memory and peripheral protection controllers); and (iii) \textit{\textbf{core}}, i.e., \textsc{uTango} boot and scheduler logic (e.g., memory and devices partition, and system timer configuration). From Table \ref{tb-loc}, it is possible to conclude that the architectural and platform-specific code represents most of the total SLoC. Since \textsc{uTango's} heavy lifting work is during boot-time, i.e., system resources partition, hardware initialization, and configuration, it is normal that these two components reflect the major part of the \textsc{uTango} code complexity ($\approx$2K SLoC). On the other side, the run-time logic, i.e., worlds scheduling and the re-partition of system resources, which is implemented in assembly, encompasses a total of $\approx$200 SLoC, corresponding to 4.6\% of the total SLoC. 

\mypara{Binary size.} To measure the size (bytes) of \textsc{uTango}, we use the GCC size tool (Berkeley format). Table \ref{tb-loc} presents the \textit{.text}, \textit{.data}, and \textit{.bss} sections, according to system component, i.e., organized by directories.
As highlighted above, target-specific functionality (e.g. SAU, SysTick, MPC, and PPC drivers) included in \textit{\textbf{arch}} and \textit{\textbf{platform}} directories represent approximately 2/3 of the total \textsc{uTango's} size. 
At boot-time, \textsc{uTango} core allocates the WCB structure and performs initialization routines. For each configured world, the system allocates 324 bytes of data for its private WCB. During WCB's initialization, \textsc{uTango} retrieves from the \textit{config} structure (60 bytes) each world's configuration. This structure is filled by the system designer to describe, per world, the memory layout, available devices, and assigned interrupts. Regarding run-time code, the total size is 488 bytes, which represents the code implementing the scheduling logic. Thus, the resulting TCB size is 4.3 KiB. 

\section{Discussion}
\label{dis}
In this section, we discuss main \textsc{uTango}'s limitations and possible future improvements.

\mypara{Secure Services.} \label{ev-sa7}
We intend to demonstrate a set of secure services, such as secure storage, cryptography operations, remote attestation, and secure updates over-the-air. These secure services will be encapsulated in dedicated NSVWs. However, depending on the peculiarities of the target platform, some hardware modules may be hardwired to the secure world. To address this challenge, we envision the development of lightweight (de-privileged) secure gateways to mediate access to secure world resources. Additional hardware primitives (i.e., secure MPU) will be leveraged to enforce isolation within the secure world.

\mypara{Non-TrustZone Platforms Support.}
As of now, \textsc{uTango} only supports Armv8-M architectures with TrustZone-M security extensions. We intend to extend this support to Armv7-M and RISC-V architectures in the future. In Armv7-M architectures, we envision \textsc{uTango} running in the highest privilege level (i.e., privileged handler mode); however, a set of challenges need to be addressed: (i) the MPU, which enforces access permissions to memory regions, has less flexibility (i.e., restricted regions sizes and alignment conditions); and (ii) the need to support trap and emulation due to special privileged instructions and imprecise bus faults \cite{Pinto2020}. 
Regarding RISC-V architectures, such challenges can be alleviated in MCUs providing machine (M), supervisor (S), and user (U) mode (although without virtual-memory support) \cite{Sa2021}. Legacy applications may run unmodified and with access to privileged operations. \textsc{uTango} will run in M-mode with full access to the hardware features, including the Physical Memory Protection (PMP) that can enforce isolation between NSVWs with higher flexibility (i.e., more fine-grained regions).

\newcommand{\yes}{\CIRCLE}
\newcommand{\no}{\Circle}
\newcommand{\half}{\LEFTcircle}
\newcommand{\unk}{?}

\begin{table*}[ht]
\centering
\caption{Examples of academic and commercial solutions that contribute with relevant TEE (or TEE-like) implementations.}
\label{tb-relw}
\scalebox{0.87}{
\begin{tabular}{llcccllcc}
\multicolumn{1}{l}{} & \multicolumn{1}{c}{\multirow{2}{*}{\textbf{Alias}}} & \multicolumn{2}{c}{\textbf{Background}} & \multicolumn{1}{c}{\multirow{2}{*}{{\textbf{Open-Source}}}} & \multicolumn{1}{c}{\multirow{2}{*}{{\textbf{Processor ISA}}}} & \multicolumn{1}{c}{\multirow{2}{*}{\parbox{2cm}{\textbf{Isolation Tech./Mech.}}}} & \multicolumn{2}{c}{\textbf{Software Architecture}} \\ 
\cline{3-4}\cline{8-9}
\multicolumn{1}{c}{} & \multicolumn{1}{c}{} & \multicolumn{1}{c}{\small\textit{Academic}} & \multicolumn{1}{c}{\small\textit{Commercial}} & \multicolumn{1}{c}{} & \multicolumn{1}{c}{} & \multicolumn{1}{c}{} & \multicolumn{1}{c}{\parbox{1.8cm}{\small{\textit{Dual-Domain}}}} & \multicolumn{1}{c}{\parbox{1.9cm}{\small{\textit{Multi-Domain}}}}  \\ 
\toprule
\multirow{8}{*}{\rotatebox{90}{\small{High-/Middle-end SoCs}}} & SeCReT \cite{SeCReT2015} & \yes & & & Armv7-A & TZ-A & \yes & \\
 & TrustICE \cite{Sun2015} & \yes & & & Armv7-A & TZ-A & & \yes \\
 & Sanctum \cite{Costan2016} & \yes & & \yes & RISC-V & PMP + HW ext. & & \yes \\
 & vTZ \cite{Hua2017} & \yes & & & Armv7-A & TZ-A & & \yes \\
 & Komodo \cite{Ferraiuolo2017} & \yes & & & Armv7-A & TZ-A & & \yes \\
 & Keystone \cite{Lee2020} & \yes & & \yes & RISC-V & PMP & & \yes \\
 & Sanctuary \cite{Brasser2019} & \yes & & & Armv8-A & TZ-A & & \yes \\
 & Timber-V \cite{Weiser2019} & \yes & & \yes & RISC-V & PMP & & \yes \\
\midrule
\midrule
\multirow{1}{*} {\rotatebox{90}{\small{Low-end MCUs}}}
 & ATF-M \cite{atfm2020} & & \yes & \yes & Armv7-M/v8-M & TZ-M/dual-core & \yes & \\
 & Kinibi-M \cite{kinibi2021} & & \yes & & Armv8-M & TZ-M & \yes & \\
 & ProvenCore-M \cite{provencore2021} & & \yes & & Armv7-M/v8-M & TZ-M/dual-core & \yes & \\
 & MultiZone \cite{Pinto2020} & & \yes & & \parbox{2.7cm}{RISC-V/Armv7-M} & PMP/MPU & & \yes \\
 & \cellcolor{black!25}\textbf{\textsc{uTango}} & \cellcolor{black!25}\yes & \cellcolor{black!25} & \cellcolor{black!25}\yes & \cellcolor{black!25}Armv8-M & \cellcolor{black!25}TZ-M & \cellcolor{black!25} & \cellcolor{black!25}\yes \\
\midrule
\midrule
\end{tabular}
}
\end{table*}

\mypara{Priority-based World Interrupt Handling.}
Despite the collected data showing a deterministic behavior (including worst-case scenarios), the current interrupt handling mechanism presents high interrupt latency in some scenarios. Nevertheless, current efforts are focused on a preemptive priority-based strategy, as mentioned in Section \ref{ui8}. When suspended, low-priority NSVW's interrupts will be temporarily configured as secure and disabled (prevent priority escalation). In contrast, high-priority NSVW's interrupts will be configured as secure and enabled; thereby, if an interrupt is triggered, the processor is preempted, and the WS will handle the execution to the high-priority NSVW. By endowing \textsc{uTango} with this mechanism, it is expected that the interrupt latency can decrease significantly; however, the implementation must be carefully crafted to prevent unacceptable overheads on the world switch time, given the increase of the WS complexity.

\section{Related Work}
\label{rw}

\par There is a rich body of runtime environments, isolation techniques and mechanisms, and architectures for secure execution and isolated environments \cite{Klein2009, Noorman2013, Koeberl2014, Brasser2015, Sun2015, Cho2016, Costan2016, Azab2016, Ferraiuolo2017, Maene2018, Hassaan2019, Brasser2019, Pinto2019_Survey, Li2019, Hahm2020, Lee2020, Bahmani2021,Hua2017, SeCReT2015, Weiser2019}. Due to the extensive list of works, we focus on three main classes of solutions: (i) TEE systems targeting high- to middle-end IoT devices, i.e., powered by application processors; (ii) TEE systems for low-end IoT devices, i.e., powered by MCUs; and reliable systems for low-end IoT devices. Table \ref{tb-relw} summarizes and compares several academic and commercial TEE systems across several dimensions, identifies solutions for low-end IoT devices (the target and niche of our work), and highlights \textsc{uTango} as the first open-source, multi-world TEE for TrustZone-M-based MCUs.



\mypara{TEEs for high-/middle-end IoT devices.}
For Arm application processors, TrustZone has been pivotal for building TEE systems. There are plenty of commercial TrustZone-assisted TEEs (e.g., QSEE, Kinibi \cite{Cerdeira2020}), but academia has been focusing on enhancing TrustZone TEE systems with increasing isolation capabilities and security guarantees. TrustICE~\cite{Sun2015} and Sanctuary~\cite{Brasser2019} leverages the TZASC to create enclaves within the normal world. TrustICE has severe multi-core limitations, while Sanctuary falls short on depending on TZASC features not available in any COTS SoC released to date. Komodo \cite{Ferraiuolo2017} strengthens software isolation between secure applications by relying on a formally verified microkernel that enables SGX-like enclaves. vTZ~\cite{Hua2017}, OSP~\cite{Cho2016}, and PrivateZone~\cite{Jang2018} leverage the hardware virtualization extensions available in the normal world to implement multiple isolated environments. TEEv~\cite{Li2019} and PrOS~\cite{Kwon20} use same-privilege techniques to secure a minimalist hypervisor in the secure world. All these systems target high- to middle-end processors and leverage techniques and mechanisms not available on resource-constrained MCUs. For RISC-V, there are two main classes of solutions. A set of works leverage the open hardware model to enhance RISC-V cores and SoCs with prime security mechanisms, e.g., Sanctum \cite{Costan2016}, HECTOR-V \cite{Nasahl2020}, TIMBER-V \cite{Weiser2019}, CURE \cite{Bahmani2021}. A second line of works leverage the standard RISC-V hardware primitives to provide frameworks for customizable TEEs, e.g., Keystone \cite{Lee2020} and MultiZone-Linux \cite{Garlati2020}.


\mypara{TEEs for MCU-powered IoT devices.}
TEE systems for resource-constrained IoT devices are in their infancy, and only a few commercial and academic solutions have been proposed so far. Janjua et al. \cite{Hassaan2019} have developed the Security MicroVisor (SuV), a pure-software TEE for resource-constrained devices that lack basic hardware-based security features such as MPU (e.g., AVR ATmega). MultiZone TEE \cite{Pinto2020} is an innovative hardware-enforced, software-defined TEE for (Armv7-M) Cortex-M and RISC-V MCUs. MultiZone leverages the Arm MPU or the RISC-V Physical Memory Protection (PMP) to create multiple isolated environments. In the context of TrustZone-M MCUs, ATF-M \cite{atfm2020} provides an open source reference implementation of a TEE for Armv8-M devices. Kinibi-M \cite{kinibi2021} and ProvenCore-M \cite{provencore2021} are preeminent examples of commercial TEE solutions adapted from existing well-established Cortex-A implementations. mTower \cite{Drozdovskyi2019} is an open source initiative from Samsung aiming at developing a TEE specially designed to protect size-constrained IoT devices based on the Cortex-M23. Contrary to the aforementioned TrustZone-M solutions, which are a strict materialization of the TrustZone dual-world architecture, \textsc{uTango} relies on a multi-world design, providing multiple isolated environments within the normal world, and thus addressing the main architectural deficiencies observed in commercial TrustZone systems while providing augmented TEE capabilities. To the best of our knowledge, MultiZone \cite{Pinto2020} is the closest solution to \textsc{uTango}. Notwithstanding, comparing to our approach, MultiZone for Arm Cortex-M requires (i) static binary translation to handle special privileged instructions and imprecise bus faults and (ii) implements trap and emulation. There is also a preeminent class of solutions that proposes a set of mechanisms for TrustZone-M devices. CoreLockr-TZ \cite{CoreLockr} is a lightweight service dispatch layer and CFI CaRE \cite{Nyman2017} implements a prime control-flow integrity (CFI) mechanism. Finally, ASSURED \cite{Asokan2018} proposes a secure firmware update framework for TrustZone-M devices.


\mypara{Reliable systems for MCU-powered IoT devices.} 
Classic approaches to provide isolation and implement reliable systems on low-end embedded devices have been evolving from constructive (language/compiler-based) memory protection \cite{Cooprider2007, Levy2017, Clements2017, Kwon2019, Peach2020} and hardware-enforced RTOS mechanisms \cite{Hahm2016, Silva2019, Hahm2020}, to lightweight virtualization infrastructures \cite{Bruns2013, Paci2016, Pan2018, Pinto2019_Voila}. Tock \cite{Levy2017} leverages limited hardware protection mechanisms as well as the type-safety features of the Rust programming language to provide a reliable multiprogramming environment for MCUs. EPOXY \cite{Clements2017} proposes a technique called privilege overlaying and uXOM \cite{Kwon2019} implements a protection mechanism that leverages the LLVM compiler to translate all memory instructions into unprivileged ones, constraining the code region using the MPU available on Cortex-M MCUs. Peach et al. proposed eWASM \cite{Peach2020}, a runtime environment to constrain memory accesses and control flow, enabled by the aWsm compiler. Several widespread embedded (RT)OSes such as Mbed OS \cite{mbed2021}, FreeRTOS \cite{freertos2021}, and Zephyr \cite{zephyr2021} have already upstream support for task isolation using the MPU. Another class of approaches have proposed lightweight virtualization solutions for resource-constrained devices.  F. Bruns et al. \cite{Bruns2013} and R. Pan et al. \cite{Pan2018} have proposed virtualization infrastructures leveraging the MPU. Pinto et al. \cite{Pinto2019_Voila} have also proposed a TrustZone-based virtualization solution for Cortex-M MCUs.

\section{Conclusion}
In this paper, we presented \textsc{uTango}, the first multi-world TEE for TrustZone-M IoT devices. Our innovative design enables the execution of multiple environments within strongly isolated compartments with increasing flexibility and security guarantees. \textsc{uTango} will be publicly available in hopes of engaging both academia and industry on research and deployment of innovative TEE solutions for the tiniest IoT devices.




\ifCLASSOPTIONcaptionsoff
  \newpage
\fi



%
\bibliographystyle{IEEEtran}
\bibliography{bib/iotj_bib}

\begin{thebibliography}{10}
\providecommand{\url}[1]{#1}
\csname url@samestyle\endcsname
\providecommand{\newblock}{\relax}
\providecommand{\bibinfo}[2]{#2}
\providecommand{\BIBentrySTDinterwordspacing}{\spaceskip=0pt\relax}
\providecommand{\BIBentryALTinterwordstretchfactor}{4}
\providecommand{\BIBentryALTinterwordspacing}{\spaceskip=\fontdimen2\font plus
\BIBentryALTinterwordstretchfactor\fontdimen3\font minus
  \fontdimen4\font\relax}
\providecommand{\BIBforeignlanguage}[2]{{%
\expandafter\ifx\csname l@#1\endcsname\relax
\typeout{** WARNING: IEEEtran.bst: No hyphenation pattern has been}%
\typeout{** loaded for the language `#1'. Using the pattern for}%
\typeout{** the default language instead.}%
\else
\language=\csname l@#1\endcsname
\fi
#2}}
\providecommand{\BIBdecl}{\relax}
\BIBdecl

\bibitem{Keoh2014}
S.~L. Keoh, S.~S. Kumar, and H.~Tschofenig, ``{Securing the Internet of Things:
  A Standardization Perspective},'' \emph{IEEE IoT-J}, 2014.

\bibitem{Alrawi2019}
O.~Alrawi, C.~Lever, M.~Antonakakis, and F.~Monrose, ``{SoK: Security
  Evaluation of Home-Based IoT Deployments},'' in \emph{Proc. of S\&P}, 2019.

\bibitem{Balzarotti2020}
E.~Cozzi, P.~Veroer, M.~Dell'Amico, Y.~Shen, L.~Bilge, and D.~Balzarotti,
  ``{The Tangled Genealogy of IoT Malware},'' in \emph{Proc. of ACSAC}, 2020.

\bibitem{Pan2018}
R.~Pan, G.~Peach, Y.~Ren, and G.~Parmer, ``Predictable virtualization on memory
  protection unit-based microcontrollers,'' in \emph{Proc. of RTAS}, 2018.

\bibitem{Pinto2019_Voila}
S.~{Pinto}, H.~{Araujo}, D.~{Oliveira}, J.~{Martins}, and A.~{Tavares},
  ``{Virtualization on TrustZone-Enabled Microcontrollers? Voilà!}'' in
  \emph{Proc. of RTAS}, 2019.

\bibitem{Sadeghi2015}
A.-R. Sadeghi, C.~Wachsmann, and M.~Waidner, ``{Security and privacy challenges
  in industrial Internet of Things},'' in \emph{Proc. of DAC}, 2015.

\bibitem{Oliveira2020}
D.~Oliveira, M.~Costa, S.~Pinto, and T.~Gomes, ``The future of low-end motes in
  the internet of things: A prospective paper,'' \emph{Electronics}, 2020.

\bibitem{Luo2020}
L.~Luo, Y.~Zhang, C.~C. Zou, X.~Shao, Z.~Ling, and X.~Fu, ``On runtime software
  security of trustzone-m based iot devices,'' \emph{CoRR}, 2020.

\bibitem{Freiling2017}
J.~Noorman, J.~V. Bulck, J.~T. M{\"{u}}hlberg, F.~Piessens, P.~Maene,
  B.~Preneel, I.~Verbauwhede, J.~G{\"{o}}tzfried, T.~M{\"{u}}ller, and F.~C.
  Freiling, ``Sancus 2.0: {A} low-cost security architecture for iot devices,''
  \emph{{ACM} Trans. Priv. Secur.}, 2017.

\bibitem{Klein2009}
G.~Klein, K.~Elphinstone, G.~Heiser, J.~Andronick, D.~Cock, P.~Derrin,
  D.~Elkaduwe, K.~Engelhardt, R.~Kolanski, M.~Norrish, T.~Sewell, H.~Tuch, and
  S.~Winwood, ``{SeL4: Formal Verification of an OS Kernel},'' in \emph{Proc.
  of SOSP}, 2009.

\bibitem{Koeberl2014}
P.~Koeberl, S.~Schulz, A.~Sadeghi, and V.~Varadharajan, ``{TrustLite: A
  Security Architecture for Tiny Embedded Devices},'' in \emph{EuroSys}, 2014.

\bibitem{Brasser2015}
F.~Brasser, B.~Mahjoub, A.~Sadeghi, C.~Wachsmann, and P.~Koeberl, ``{TyTAN:
  Tiny trust anchor for tiny devices},'' in \emph{Proc. of DAC}, 2015.

\bibitem{Sun2015}
H.~Sun, K.~Sun, Y.~Wang, J.~Jing, and H.~Wang, ``{Trustice: Hardware-assisted
  isolated computing environments on mobile devices},'' in \emph{Proc. of DSN},
  2015.

\bibitem{Costan2016}
V.~Costan, I.~Lebedev, and S.~Devadas, ``{Sanctum: Minimal Hardware Extensions
  for Strong Software Isolation},'' in \emph{Proc. of {USENIX}}, 2016.

\bibitem{Ferraiuolo2017}
A.~Ferraiuolo, A.~Baumann, C.~Hawblitzel, and B.~Parno, ``Komodo: Using
  verification to disentangle secure-enclave hardware from software,'' in
  \emph{Proc. of SOSP}, 2017.

\bibitem{Maene2018}
P.~Maene, J.~Götzfried, R.~de~Clercq, T.~Müller, F.~Freiling, and
  I.~Verbauwhede, ``{Hardware-Based Trusted Computing Architectures for
  Isolation and Attestation},'' \emph{IEEE Trans. on Computers}, 2018.

\bibitem{Hassaan2019}
H.~Janjua, M.~Ammar, B.~Crispo, and D.~Hughes, ``{Towards a Standards-Compliant
  Pure-Software Trusted Execution Environment for Resource-Constrained Embedded
  Devices},'' in \emph{Proc. of SysTEX}, 2019.

\bibitem{Brasser2019}
F.~Brasser, D.~Gens, P.~Jauernig, A.-R. Sadeghi, and E.~Stapf, ``{SANCTUARY:
  ARMing TrustZone with User-space Enclaves},'' in \emph{Proc. of NDSS}, 2019.

\bibitem{Pinto2019_Survey}
S.~Pinto and N.~Santos, ``{Demystifying Arm TrustZone: A Comprehensive
  Survey},'' \emph{ACM Comput. Surv.}, 2019.

\bibitem{Li2019}
W.~Li, Y.~Xia, L.~Lu, H.~Chen, and B.~Zang, ``{TEEv: Virtualizing Trusted
  Execution Environments on Mobile Platforms},'' in \emph{Proc. of VEE}, 2019.

\bibitem{Hahm2020}
S.~il~Hahm, J.~Kim, A.~Jeong, H.~Yi, S.~Chang, K.~SN, A.~Chauhan, and S.~P.
  Cherian, ``{Reliable Real-Time Operating System for IoT Devices},''
  \emph{IEEE IoT-J}, 2020.

\bibitem{Lee2020}
D.~Lee, D.~Kohlbrenner, S.~Shinde, K.~Asanovi\'{c}, and D.~Song, ``{Keystone:
  An Open Framework for Architecting Trusted Execution Environments},'' in
  \emph{Proc. of EuroSys}, 2020.

\bibitem{Bahmani2021}
R.~Bahmani, F.~Brasser, G.~Dessouky, P.~Jauernig, M.~Klimmek, A.-R. Sadeghi,
  and E.~Stapf, ``{{CURE}: A Security Architecture with CUstomizable and
  Resilient Enclaves},'' in \emph{Proc. of {USENIX} Security}, 2021.

\bibitem{Hua2017}
Z.~Hua, J.~Gu, Y.~Xia, H.~Chen, B.~Zang, and H.~Guan, ``{vTZ: Virtualizing ARM
  TrustZone},'' in \emph{Proc. of {USENIX} Security}, 2017.

\bibitem{Pinto2020}
S.~Pinto and C.~Garlati, ``{Multi Zone Security for Arm Cortex-M Devices},'' in
  \emph{Proc. of Embedded World Conf.}, 2020.

\bibitem{Nyman2017}
T.~Nyman, J.~Ekberg, L.~Davi, and N.~Asokan, ``{CFI CaRE: Hardware-Supported
  Call and Return Enforcement for Commercial Microcontrollers},'' in
  \emph{Proc. of RAID}, 2017.

\bibitem{Asokan2018}
N.~Asokan, T.~Nyman, N.~Rattanavipanon, A.-R. Sadeghi, and G.~Tsudik,
  ``{ASSURED: Architecture for Secure Software Update of Realistic Embedded
  Devices},'' \emph{IEEE TCAD}, 2018.

\bibitem{Zhang2016}
N.~Zhang, H.~Sun, K.~Sun, W.~Lou, and Y.~T. Hou, ``{CacheKit: Evading Memory
  Introspection Using Cache Incoherence},'' in \emph{Proc. of S\&P}, 2016.

\bibitem{Tang2017}
A.~Tang, S.~Sethumadhavan, and S.~Stolfo, ``{CLKSCREW}: Exposing the perils of
  security-oblivious energy management,'' in \emph{Proc. of {USENIX} Security},
  2017.

\bibitem{Keegan2019}
K.~Ryan, ``{Hardware-Backed Heist: Extracting ECDSA Keys from Qualcomm's
  TrustZone},'' in \emph{Proc. of CCS}, 2019.

\bibitem{Cerdeira2020}
D.~{Cerdeira}, N.~{Santos}, P.~{Fonseca}, and S.~{Pinto}, ``{SoK: Understanding
  the Prevailing Security Vulnerabilities in TrustZone-assisted TEE Systems},''
  in \emph{Proc. of S\&P}, 2020.

\bibitem{atfm2020}
Arm, ``{Arm Trusted Firmware},'' \url{www.trustedfirmware.org/}, accessed:
  2021-01-12.

\bibitem{kinibi2021}
Trustonic, ``{Trustonic Kinibi-M},'' \url{www.trustonic.com/technology/},
  accessed: 2021-01-12.

\bibitem{Arm2018}
Arm, ``{Arm TrustZone technology for the ARMv8-M architecture},'' Arm Ltd.,
  Tech. Rep., Oct 2018.

\bibitem{Arm2019}
Arm, ``{PSA Application Guide},'' Arm Ltd., Tech. Rep., 2019.

\bibitem{ArmKeil2019}
Arm, ``{Using TrustZone on Armv8-M},'' Arm Ltd., Tech. Rep., Sep 2019.

\bibitem{Cho2016}
Y.~Cho, J.~Shin, D.~Kwon, M.~Ham, Y.~Kim, and Y.~Paek, ``{Hardware-Assisted
  On-Demand Hypervisor Activation for Efficient Security Critical Code
  Execution on Mobile Devices},'' in \emph{Proc. of {USENIX} ATC}, 2016.

\bibitem{Jang2018}
J.~Jang, C.~Choi, J.~Lee, N.~Kwak, S.~Lee, Y.~Choi, and B.~B. Kang,
  ``{PrivateZone: Providing a Private Execution Environment Using ARM
  TrustZone},'' \emph{IEEE Trans. on Dependable and Secure Computing}, 2018.

\bibitem{Garlati2020}
C.~Garlati and S.~Pinto, ``{A Clean Slate Approach to Linux Security RISC-V
  Enclaves},'' in \emph{Proc. of Embedded World Conf.}, 2020.

\bibitem{Pinto2017}
S.~Pinto, J.~Pereira, T.~Gomes, A.~Tavares, and J.~Cabral, ``Ltzvisor:
  Trustzone is the key,'' in \emph{ECRTS}, 2017.

\bibitem{Arm2020}
Arm, ``{Armv8-M Secure Stack Sealing},'' Arm Ltd., Tech. Rep., 2020.

\bibitem{armpsa2021}
Arm, ``{Arm PSA},''
  \url{www.arm.com/why-arm/architecture/platform-security-architecture},
  accessed: 2021-01-30.

\bibitem{atfm2022}
Arm, ``{ATF-M Security Vulnerabilities},''
  \url{https://tf-m-user-guide.trustedfirmware.org/docs/security/security_advisories/index.html},
  accessed: 2021-01-12.

\bibitem{Saltzer1975}
J.~Saltzer and M.~Schroeder, ``The protection of information in computer
  systems,'' \emph{Proc. of IEEE}, 1975.

\bibitem{ArmPSA2019}
Arm, ``{PSA Firmware Framework},'' Arm Ltd., Tech. Rep., 2019.

\bibitem{Bagchi2019}
N.~S. Almakhdhub, A.~A. Clements, M.~Payer, and S.~Bagchi, ``{BenchIoT: A
  Security Benchmark for the Internet of Things},'' in \emph{Proc. of DSN},
  2019.

\bibitem{embench20}
D.~Patterson, J.~Bennett, P.~Dabbelt, C.~Garlati, G.~S. Madhusudan, and
  T.~Mudge, ``{Embench: A Modern Embedded Benchmark Suite},'' www.embench.org/,
  accessed: 2020-12-22.

\bibitem{Sa2021}
B.~Sa, J.~Martins, and S.~Pinto, ``{A First Look at RISC-V Virtualization from
  an Embedded Systems Perspective},'' \emph{IEEE Transactions on Computers},
  no.~01, pp. 1--1, nov 2021.

\bibitem{SeCReT2015}
J.~S. Jang, S.~Kong, M.~Kim, D.~Kim, and B.~B. Kang, ``{SeCReT: Secure Channel
  between Rich Execution Environment and Trusted Execution Environment},'' in
  \emph{Proc. of NDSS}, 2015.

\bibitem{Weiser2019}
S.~Weiser, M.~Werner, F.~Brasser, M.~Malenko, S.~Mangard, and A.~Sadeghi,
  ``{TIMBER-V:} tag-isolated memory bringing fine-grained enclaves to
  {RISC-V},'' in \emph{Proc. of NDSS}, 2019.

\bibitem{provencore2021}
Prove\&Run, ``{ProvenCore-M},'' \url{www.provenrun.com/products/provencore/},
  accessed: 2021-01-12.

\bibitem{Noorman2013}
J.~Noorman, P.~Agten, W.~Daniels, R.~Strackx, A.~V. Herrewege, C.~Huygens,
  B.~Preneel, I.~Verbauwhede, and F.~Piessens, ``Sancus: Low-cost trustworthy
  extensible networked devices with a zero-software trusted computing base,''
  in \emph{Proc. of {USENIX} Security}, 2013.

\bibitem{Azab2016}
A.~M. Azab, K.~Swidowski, R.~Bhutkar, J.~Ma, W.~Shen, R.~Wang, and P.~Ning,
  ``{SKEE: A Lightweight Secure Kernel-level Execution Environment for ARM},''
  in \emph{Proc. of NDSS}, 2016.

\bibitem{Kwon20}
D.~Kwon, J.~Seo, Y.~Cho, B.~Lee, and Y.~Paek, ``{PrOS: Light-Weight Privatized
  Se cure OSes in ARM TrustZone},'' \emph{IEEE Trans. on Mobile Computing},
  2020.

\bibitem{Nasahl2020}
P.~Nasahl, R.~Schilling, M.~Werner, and S.~Mangard, ``{{HECTOR-V:} {A}
  Heterogeneous {CPU} Architecture for a Secure {RISC-V} Execution
  Environment},'' \emph{CoRR}, 2020.

\bibitem{Drozdovskyi2019}
T.~A. Drozdovskyi and O.~S. Moliavko, ``{mTower: Trusted Execution Environment
  for MCU-based devices},'' \emph{JOSS}, 2019.

\bibitem{CoreLockr}
{Sequitur Labs}, ``{CoreLockr-TZ},''
  \url{https://www.sequithowpublishedabs.com/corelockrtz/}, 2017, accessed:
  2021-01-01.

\bibitem{Cooprider2007}
N.~Cooprider, W.~Archer, E.~Eide, D.~Gay, and J.~Regehr, ``{Efficient Memory
  Safety for TinyOS},'' in \emph{Proc. of SenSys}, 2007.

\bibitem{Levy2017}
A.~Levy, B.~Campbell, B.~Ghena, D.~B. Giffin, P.~Pannuto, P.~Dutta, and
  P.~Levis, ``{Multiprogramming a 64kB Computer Safely and Efficiently},'' in
  \emph{Proc. of SOSP}, 2017.

\bibitem{Clements2017}
A.~A. Clements, N.~S. Almakhdhub, K.~S. Saab, P.~Srivastava, J.~Koo, S.~Bagchi,
  and M.~{Payer}, ``{Protecting Bare-Metal Embedded Systems with Privilege
  Overlays},'' in \emph{Proc. of S\&P}, 2017.

\bibitem{Kwon2019}
D.~Kwon, J.~Shin, G.~Kim, B.~Lee, Y.~Cho, and Y.~Paek, ``{uXOM: Efficient
  eXecute-Only Memory on {ARM} Cortex-M},'' in \emph{Proc. of {USENIX}
  Security}, 2019.

\bibitem{Peach2020}
G.~Peach, R.~Pan, Z.~Wu, G.~Parmer, C.~Haster, and L.~Cherkasova, ``{eWASM:
  Practical Software Fault Isolation for Reliable Embedded Devices},''
  \emph{IEEE Trans. on Computer-Aided Design of Integrated Circuits and
  Systems}, 2020.

\bibitem{Hahm2016}
O.~Hahm, E.~Baccelli, H.~Petersen, and N.~Tsiftes, ``{Operating Systems for
  Low-End Devices in the Internet of Things: A Survey},'' \emph{IEEE IoT-J},
  2016.

\bibitem{Silva2019}
M.~{Silva}, D.~{Cerdeira}, S.~{Pinto}, and T.~{Gomes}, ``{Operating Systems for
  Internet of Things Low-End Devices: Analysis and Benchmarking},'' \emph{IEEE
  IoT-J}, 2019.

\bibitem{Bruns2013}
F.~Bruns, D.~Kuschnerus, and A.~Bilgic, ``{Virtualization for Safety-critical,
  Deeply-embedded Devices},'' in \emph{Proc. of SAC}, 2013.

\bibitem{Paci2016}
F.~Paci, D.~Brunelli, and L.~Benini, ``{Lightweight IO virtualization on MPU
  enabled microcontrollers},'' \emph{ACM SIGBED Review}, 2018.

\bibitem{mbed2021}
Arm, ``{Mbed OS},'' \url{https://os.mbed.com/mbed-os/}, accessed: 2021-01-12.

\bibitem{freertos2021}
Amazon, ``{FreeRTOS},'' \url{www.freertos.org/}, accessed: 2021-01-12.

\bibitem{zephyr2021}
L.~F. Project, ``{Zephyr},'' \url{https://zephyrproject.org}, accessed:
  2021-01-12.

\end{thebibliography}

%

\vskip -35pt
\baselineskip -35pt
\begin{IEEEbiography}[{\includegraphics[width=1in,height=1.25in,clip,keepaspectratio]{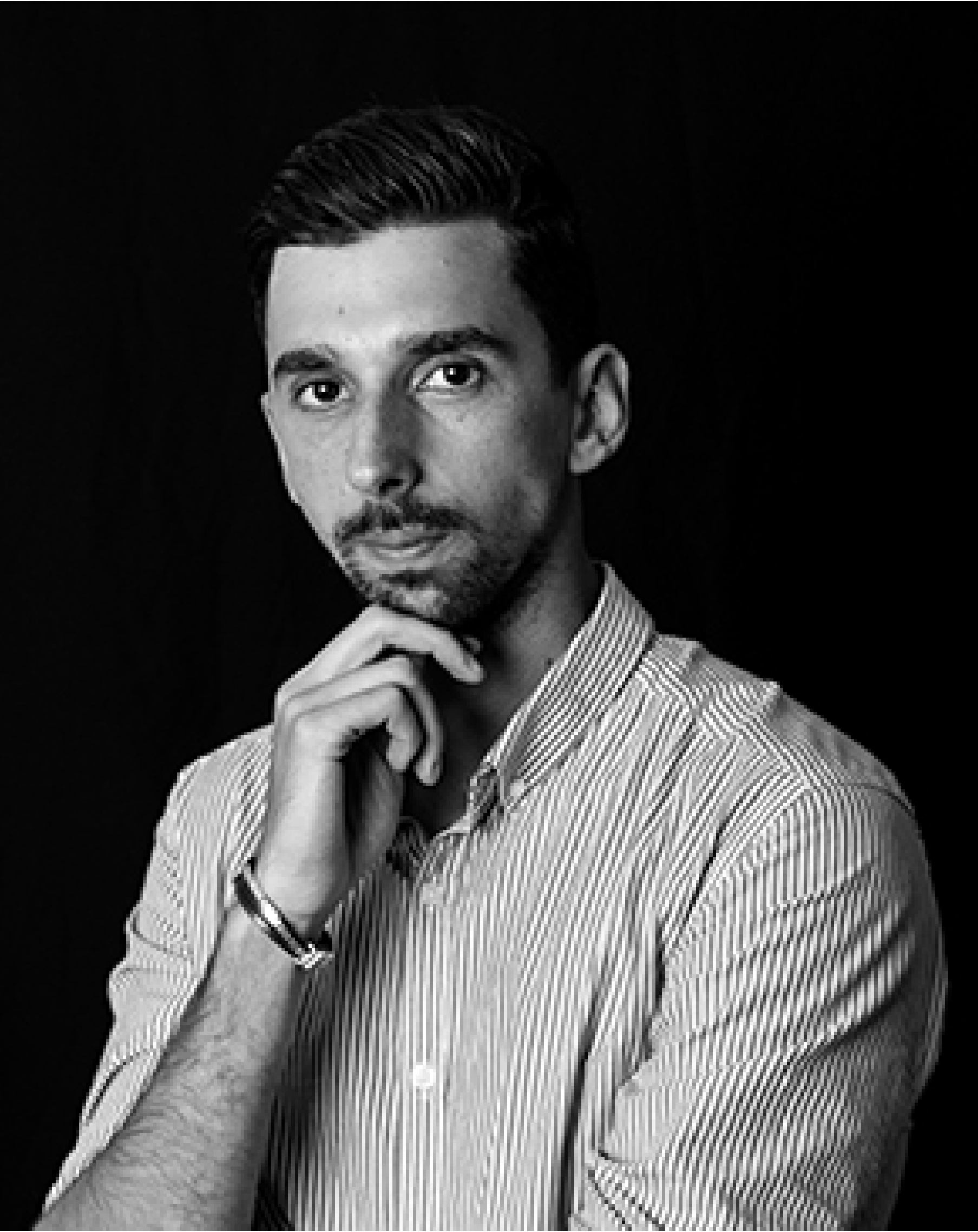}}]{Daniel Oliveira}
is an R\&D Embedded Systems Engineer, and more recently a PhD Candidate, with a strong background in (real-time) operating systems, embedded security, as well as in automotive HMI systems. He holds an MSc in Electronics and Computer Engineering from University of Minho (Portugal) where he exploited Arm TrustZone technology. His main research interests include virtualization, real-time operating systems, and security for low-end embedded devices. Contact him at daniel.oliveira@dei.uminho.pt 
\end{IEEEbiography}

\vskip -35pt
\baselineskip -35pt
\begin{IEEEbiography}[{\includegraphics[width=1in,height=1.25in,clip,keepaspectratio]{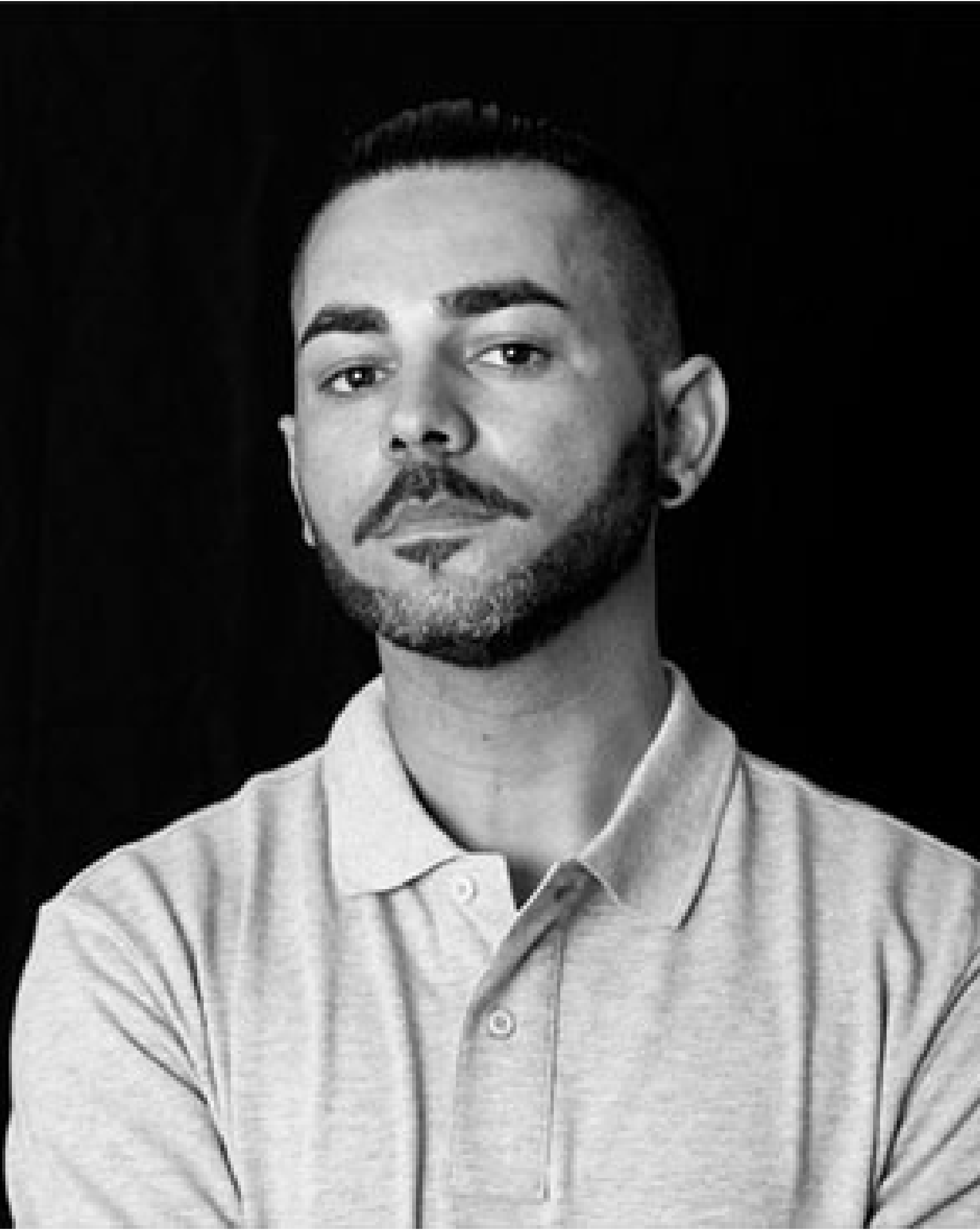}}]{Tiago Gomes}
has received the master's degree in Telecommunications Engineering and Ph.D. degree in Electronics and Computers Engineering from the University of Minho, Portugal. He is a Research Scientist and Invited Professor at the University of Minho. His current research interests include embedded systems hardware/software co-design for resource constrained wireless devices, wireless protocols for low-rate wireless personal area networks and network protocols for the Internet of Things low-end devices. Contact him at mr.gomes@dei.uminho.pt.
\end{IEEEbiography}

\vskip -35pt
\baselineskip -35pt
\begin{IEEEbiography}[{\includegraphics[width=1in,height=1.25in,clip,keepaspectratio]{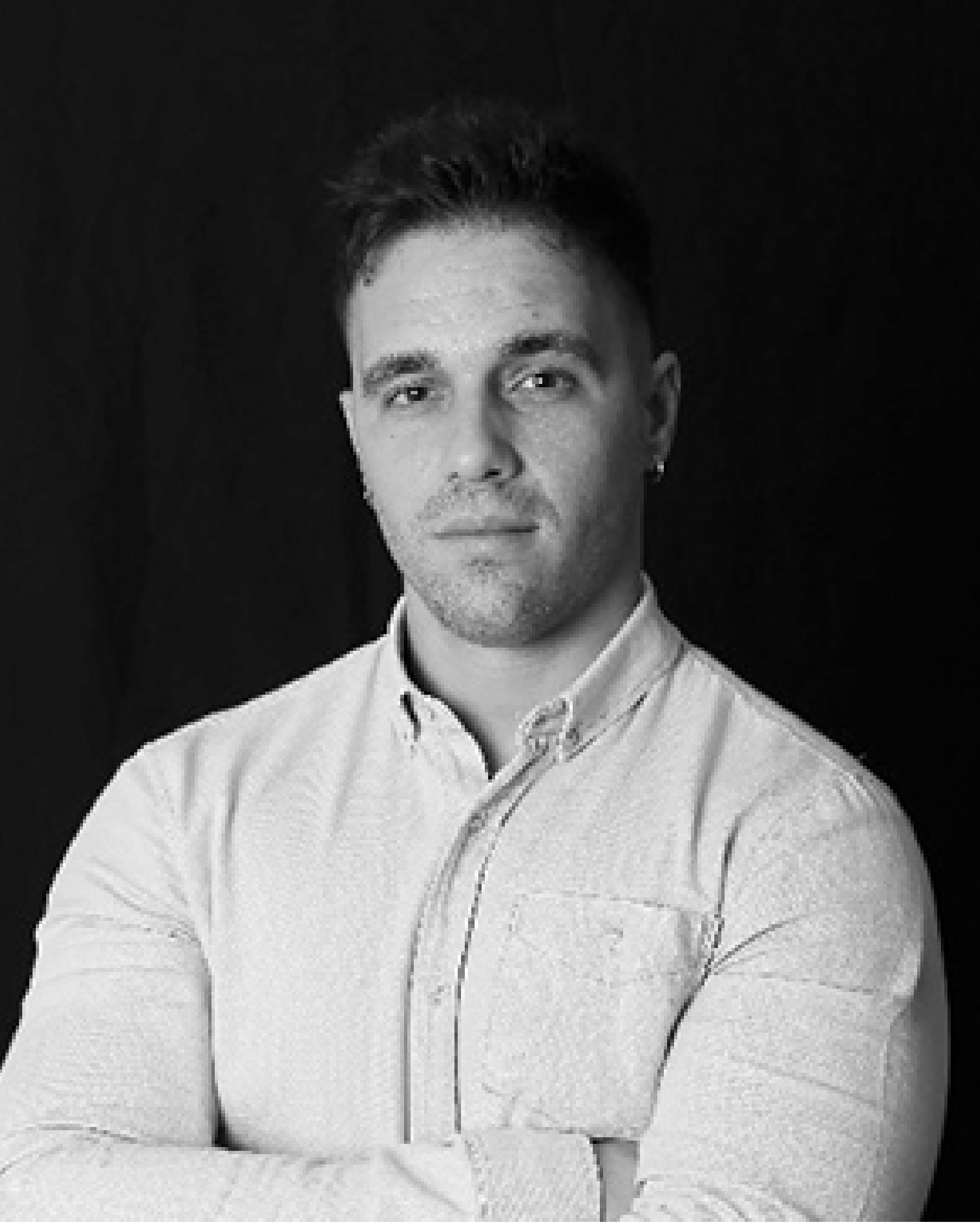}}]{Sandro Pinto}
is an Associate Research Professor at the University of Minho, Portugal. He holds a Ph.D. in Electronics and Computer Engineering. During his Ph.D., Sandro was a visiting researcher at the Asian Institute of Technology (Thailand), University of Wurzburg (Germany), and Jilin University (China). Sandro has a deep academic background and several years of industry collaboration focusing on operating systems, virtualization, and security for embedded, cyber-physical, and IoT-based systems. He has published dozens of scientific papers in top-tier conferences/journals. Contact him at sandro.pinto@dei.uminho.pt.
\end{IEEEbiography}




\end{document}